\documentclass[usenatbib]{mn2e}
\usepackage{graphicx}
\usepackage{verbatim}
\usepackage{footmisc}
   \title[Photometry of cometary nuclei]{Photometry of cometary nuclei: Rotation rates, colours and a comparison with Kuiper Belt Objects\thanks{Based on observations collected at the European Southern Observatory, Chile. Proposal: ESO No. 74.C-0125.} 
   }

   \author[C. Snodgrass et al.]{C. Snodgrass$^{1,2}$\thanks{E-mail: csnodgra@eso.org}, S. C. Lowry$^1$ and A. Fitzsimmons$^1$\\
   $^1$Astrophysics Research Centre, School of Physics and Astronomy, Queen's University Belfast, Belfast BT7 1NN, UK\\
   $^2$European Southern Observatory, Alonso de C\'ordova 3107, Vitacura, Santiago, Chile
        }

\begin{document}

   \date{Received <date> / Accepted <date>}

   \maketitle

   \begin{abstract}
We present time-series data on Jupiter Family Comets (JFCs) 17P/Holmes, 47P/Ashbrook-Jackson and 137P/Shoemaker-Levy 2. In addition we also present results from `snap-shot' observations of comets 43P/Wolf-Harrington, 44P/Reinmuth 2, 103P/Hartley 2 and 104P/Kowal 2 taken during the same run. The comets were at heliocentric distances of between 3 and 7 AU at this time. We present measurements of size and activity levels for the snap-shot targets. The time-series data allow us to constrain rotation periods and shapes, and thus bulk densities. We also measure colour indices $(V-R)$ and $(R-I)$ and reliable radii for these comets. We compare all of our findings to date with similar results for other comets and Kuiper Belt Objects (KBOs). We find that the rotational properties of nuclei and KBOs are very similar, that there is evidence for a cut-off in bulk densities at $\sim 0.6$ g cm$^{-3}$ in both populations, and the colours of the two populations show similar correlations. For JFCs there is no observational evidence for the optical colours being dependant on either position in the orbit or on orbital parameters.
\end{abstract}

 
%

\section{Introduction}

It was only in the 1950s that the modern description of comets, as bodies with a small nucleus of ice and rock that produce a coma through sublimation, and tails by solar wind pressure on this coma, was postulated \citep{Whipple50,Whipple51}. Furthermore, it is difficult to study the nucleus itself, as comets are generally active (outgassing to produce a coma) when near to the Earth, and when active the flux reflected by the coma greatly exceeds the flux from the nucleus.

There are three techniques by which this difficulty can be overcome at optical wavelengths. First, the nucleus can be seen within the coma by employing space-craft to image the comet from very close range. This was achieved with great success by the {\it VEGA} and {\it Giotto} missions at comet 1P/Halley during 1986 \citep{Sagdeev86,Keller86}, proving Whipple's `dirty snow-ball' model, and by the recent missions {\it Deep Space 1} at 19P/Borrelly \citep{Soderblom02}, {\it Stardust} at 81P/Wild 2 \citep{Brownlee04} and {\it Deep Impact} at 9P/Temple 1, the last of these making contact with the nucleus in spectacular fashion \citep{Ahearn05}. The {\it Rosetta} space-craft is on its way to 67P/Churyumov-Gerasimenko \citep{Schwehm99}, and will map its nucleus in detail and land a probe on the surface. While these missions provide the most complete information on the nucleus, they are expensive and limited in the number of targets they can observe. 

A second method relies not on going to the comet, but waiting for the comet to come to us. By taking advantage of very close passes of comets to the Earth, and very high resolution imaging using the Hubble Space Telescope (HST), the signal from the nucleus can be separated from the surrounding coma \citep[e.g.][]{Lamy99}. This method can only be applied to a limited number of comets, due to the rarity of close passes of comets to the Earth.

The third method, which we employ here, and have previously applied to comets 7P/Pons-Winnecke, 14P/Wolf and 92P/Sanguin \citep*{Snodgrass05}, is to image comets when inactive at large distances. Our programme of obtaining nuclear properties has concentrated on Jupiter Family Comets (JFCs), which are short period comets whose orbital evolution is dominated by the planet Jupiter. Believed to come from the Kuiper Belt \citep{Ip+Fernandez97}, these nuclei are left over remnants from the formation of the solar system. JFCs are generally found to be active when at heliocentric distances of $R_{\rm h} \le 3$ AU, as within this distance the sublimation rate for water ice (the primary volatile constituent of nuclei) increases dramatically. This method is difficult due to the small size (radii of a few kilometres) and very low albedos (typically $\sim$ 4\%) of nuclei, which makes them very faint targets at large distances. However, by careful selection of targets with good previous absolute magnitude estimates (i.e. the compilation by \citealt{Lamy-chapter}) that have $R_{\rm h} > 3$ AU, we can find inactive nuclei which are observable at sufficient $S/N$ using 4m class telescopes.

We employ time-series photometry to constrain the shape and bulk density of nuclei; taking data over a series of days to find the rotation period from the periodic variations in the brightness of the nucleus. Additionally, `snap-shot' observations only give an effective radius at one generally unknown rotational phase. The time-series data allows the true effective radius to be measured. Here we present time-series photometry of three JFC nuclei, in addition to results from `snap-shot' observations of four others.

In section~\ref{obs} we describe the observations and data reduction procedures employed in this work. We explain our analysis methods, and present our results from the snap-shot and time-series data, in sections \ref{ss-results} and \ref{lc-results}. Section~\ref{discussion} discusses our findings in terms of results for other comets, draws conclusions about the population of JFCs as a whole, and compares JFCs with Kuiper Belt Objects (KBOs), the supposed parent population for JFCs. Finally, section~\ref{summary} summarises all of our results.


\section{Observations and data reduction}\label{obs}

We used the 3.6m New Technology Telescope (NTT) at the European Southern Observatory's La Silla site on the nights of the 5th to 7th March 2005. The later half of the 1st night and the third night were lost to poor weather, but conditions were photometric during the remaining time. The comets targeted were at heliocentric distances between 3 and 7 AU; table~\ref{observations} gives the orbital positions of the comets at this time and a summary of the exposures taken. CCD imaging was performed using the red arm of the EMMI instrument which is mounted at the f/11 Naysmith-B focus of the NTT. The EMMI red arm contains a mosaic of 2 MIT/LL 2048$\times$4096 CCDs, and was used in 2$\times$2 binning mode to give a pixel scale of 0.332 arcsec per pixel. The effective field of view is 9.1$\times$9.9 arcmin$^2$. The images were taken primarily through the Bessell $R$ band filter, with at least one set of images taken through the Bessell $V$ and $I$ filters each night for each comet to allow measurement of colour indices. All images were taken with the telescope tracking at the sidereal rate, with exposure times chosen so that the apparent motion of the comet would be less than 0.5\arcsec{}, and would thus remain within the seeing disk. This allowed us to perform accurate measurement of the stellar-background point spread function (PSF), and highly accurate differential photometry. The $FWHM$ of the stellar-background PSF was found to vary between 0.6\arcsec{} and 3.1\arcsec{} over the two nights, with a median of 1.1\arcsec. The $VRI$-filter exposure times used are listed in table~\ref{observations}. We observed each comet in blocks of exposures lasting $\sim$ 20 minutes, and cycled between each of the comets that were visible at any given point in the night. This procedure gave us good temporal coverage without losing too much time slewing the telescope between targets.

The reduction was performed using standard IRAF tasks  \citep{IRAF1,IRAF2}. Bias-subtraction was performed utilising the prescan strip of CCD 1. Twilight sky exposures were acquired during the evening and morning, and these were combined to produce a flat-field image for each filter for each night. We have previously \citep{Snodgrass05} found that there is a variation in illumination of NTT flat-field images with rotator angle, due to scattered light from the Naysmith mirror baffle. To allow for this, we took each set of flats at a number of rotator angles, which when averaged correct the problem.

Differential photometry was performed on the comets using the IRAF packages DIGIPHOT and APPHOT \citep{IRAF3}. Aperture radii were set equal (to the nearest integer pixel) to the $FWHM$ of the stellar PSF, which has previously been found to be the optimum for maximising $S/N$ \citep{Howell89}. Choosing aperture radii as a function of stellar PSF also counters the problem of variations in measured brightness with changing seeing, which is encountered when using small fixed apertures \citep{Licandro00}. In addition, photometry of the field stars used to measure the differential magnitudes, and of standard stars from the \citet{Landolt} catalogue, was performed using an aperture of diameter 10\arcsec. The smaller aperture was used for differential photometry between the comets and field stars, while the larger diameter was used for photometric calibration and was that used by \citet{Landolt}. Using the IRAF package PHOTCAL these standard star measurements gave the zero-point, extinction coefficient and colour term for each filter for each night. These were then used to calculate the magnitudes of the field stars in each frame, taking the mean of these values gave us a very accurate measurement of the brightness of our comparison stars. Adding this value to each of the differential comet magnitudes gave us accurate calibrated comet magnitudes. There is also a term including the difference in colour between the comet and stars involved in the transformation from differential to calibrated magnitudes, but this is negligible when the comparison stars and the comet have similar colours. Care was taken to select comparison stars with colours in the range expected for cometary nuclei to minimise any uncertainties in this colour term, although experimentation found this method of obtaining calibrated photometry to be robust and independent of which comparison stars were chosen. 

\begin{table*}
\begin{minipage}[t]{2\columnwidth}

\caption{Summary of observations obtained using NTT+EMMI.}             
\label{observations}      
\centering                         
\begin{tabular}{@{}lcccccc@{}}        
\hline              
Comet & UT Date & $R_\mathrm{h}$ [AU]\footnote{Superscripts $I$ and $O$ refer to whether the comet is inbound (pre-perihelion) or outbound (post-perihelion). The variations in $R_\mathrm{h}$, $\Delta$ and $\alpha$ over the course of any one night were smaller than the significance quoted here.} & $\Delta$ [AU] & $\alpha$ [deg.] & $N_{\rm frames}\times$Filter & Exposure time [s]\\    
\hline                        
17P/Holmes & 05/03/2005 & 4.66$^I$ & 3.93 & 9.0 & 14$\times$$R$, 2$\times$$V$, 2$\times$$I$ & 100\\
& 06/03/2005 & 4.66$^I$ & 3.91 & 8.9 & 25$\times$$R$, 3$\times$$V$, 3$\times$$I$ & 100\\

43P/Wolf-Harrington & 05/03/2005 & 3.30$^O$ & 2.61 & 13.9 & 4$\times$$R$ & 100 \\

44P/Reinmuth 2 & 06/03/2005 & 5.17$^I$ & 4.21 & 3.0 & 2$\times$$R$ & 80 \\

47P/Ashbrook-Jackson & 05/03/2005 & 5.42$^I$ & 4.48 & 3.5 & 19$\times$$R$, 2$\times$$V$, 2$\times$$I$ & 85\\ 
& 06/03/2005 & 5.42$^I$ & 4.47 & 3.2 & 34$\times$$R$, 2$\times$$V$, 2$\times$$I$ & 85\\ 

103P/Hartley 2 & 06/03/2005 & 3.24$^O$ & 2.28 & 5.0 & 3$\times$$R$ & 45 \\

104P/Kowal 2 & 05/03/2005 & 3.06$^O$ &  2.31 & 13.8 & 2$\times$$R$ & 80 \\

137P/Shoemaker-Levy 2 & 06/03/2005 & 6.95$^I$ & 6.17 & 5.3 & 22$\times$$R$, 2$\times$$V$, 2$\times$$I$ & 140\\ 

\hline                                   
\end{tabular}
\medskip
\end{minipage}
\end{table*}


\section{Results from snap-shot imaging}
\label{ss-results}

We first obtained `snap-shot' images of all our potential target comets, to ascertain their suitability for time-series photometry. For four comets which were not selected as primary targets, we were still able to obtain measurements or limits on the brightness and activity level of the nuclei from these data. These results are summarised in table~\ref{results_snapshots}. 

\begin{table*}
\begin{minipage}[t]{2\columnwidth}
\caption{Derived values and limits on radii and activity from snap-shot photometry on four comets.}             
\label{results_snapshots}      
\centering                         
\begin{tabular}{@{}lccccc@{}}        
\hline                 
Comet & $m_R$ & $m_c$ & $m_R(1,1,0)$ & $r_{\mathrm{N}}$ [km] & $Af\rho$ [cm]\footnote{$Af\rho$ measured through an aperture of radius 5\arcsec.}  \\    
\hline
{\bf UNDETECTED}\\
104P &  $\ge$23.3\footnote{3$\sigma$ limiting magnitudes.\label{fn2}}  & - & $\ge$18.5\footref{fn2} & $\le$0.56\footnote{3$\sigma$ upper limits.\label{fn3}} & $\le$5.9\footref{fn3} \\
{\bf UNRESOLVED}\\
44P & 22.62$\pm$0.12 & $\ge$26.9 & 15.83$\pm$0.12 & 1.96$\pm$0.11 & $\le$40.4\footref{fn3}\\
{\bf ACTIVE}\\
43P & 20.53$\pm$0.03 & - & $\ge$15.37 & $\le$2.42 & 206$\pm$2 \\
103P & 18.72$\pm$0.02 & - & $\ge$14.21 & $\le$4.13 & 196$\pm$1 \\
\hline                                   
\end{tabular}
\medskip
\end{minipage}
\end{table*}

The measured $R$-band magnitude $m_R$ is related to the effective radius of the nucleus $r_{\mathrm{N}}$ (m) by \citep{Russell}
\begin{equation}\label{rneqn}
A_R r^2_{\mathrm{N}} = 2.238 \times 10^{22} R^2_h \Delta^2 10^{0.4(m_\odot - m_R + \beta\alpha)}
\end{equation}
where $A_R$ is the geometric albedo and $m_\odot=-27.09$ is the apparent magnitude of the Sun, both in the $R$-band. $R_h$ and $\Delta$ are the heliocentric and geocentric distances in AU, and $\alpha$ and $\beta$ are the phase angle in degrees and the phase coefficient in magnitudes per degree. Without data acquired over a range of phase angles, we must assume a value for the phase coefficient. We take the commonly assumed value of $\beta$ = 0.035 mag.~deg$^{-1}$ \citep{Lamy-chapter}. In the absence of simultaneous thermal observations, the albedo must also be assumed. Comets have been found to be exceptionally dark bodies; we take the typical value of $A_R$ = 0.04 \citep{Lamy-chapter}. We also give the absolute magnitude of each nucleus, which allows comparisons without having to assume an albedo. This is given by 
\begin{equation}\label{r110eqn}
m_R(1,1,0) = m_R - 5\log(R_h \Delta) - \beta\alpha,
\end{equation}
and is the magnitude that would be measured at a hypothetical point at unit heliocentric and geocentric distances, and at $\alpha$ = 0\degr. 

\subsection{Undetected comets}

\citet*{Lowry03} have previously observed 104P/Kowal~2 when inactive at a heliocentric distance of 3.9 AU, in snap-shot data acquired with the 1.0m Jacobus Kapteyn Telescope (JKT) on La Palma in June 1999. They obtained a radius measurement of 1.0$\pm$0.5 km, again assuming an albedo of 0.04, which allowed us to predict an apparent nuclear magnitude of 21.1 during our NTT observations, making this one of our primary targets. However, this comet was not immediately detected in our initial snap-shot imaging and removed from the target list for time-series observations. Upon subsequent processing it became clear that the comet was undetectable even in images combining both $R$-band frames. The comet has a well known orbit, having been observed during four apparitions since its discovery in 1979 and most recently in February 2004, and an astrometric fit allowed us to predict the precise position of the comet at the time of observation. There was no object detected to a $m_R =24.65\pm0.46$ within the expected area. Using equation~\ref{r110eqn} this gives a 3$\sigma$ limiting magnitude of $m_R(1,1,0) \ge$ 18.53, while equation~\ref{rneqn} allows us to put a 3$\sigma$ upper limit on the radius of the nucleus of 0.56 km, assuming $A_R$ = 0.04.

This limit is in agreement with the previous measurement, although at the small end of the expected size. Comets are known to spontaneously disrupt \citep{Boehnhardt04}; break-up of this nucleus during its perihelion passage in May 2004 would also explain our non-detection. However Reach et al.\footnote{A Google search found the 104P observations on Yan Fernandez's useful page of the {\it Spitzer} logs for Solar System objects - {\tt http://www.physics.ucf.edu/$\sim$yfernandez/sss.html}}~observed the comet with the {\it Spitzer} infra-red space telescope at almost the same time as our NTT observations. The comet was detected at the expected position in the {\it Spitzer} data (Reach, private communication), so it has not broken up and therefore our non-detection gives a true upper limit to the size.

\subsection{Unresolved comets}

   \begin{figure}
   \centering
   \includegraphics[width=0.47\textwidth]{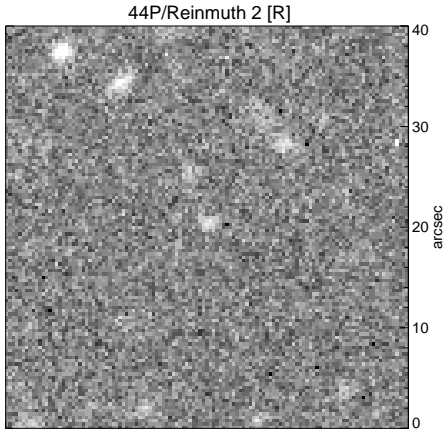}
      \caption{Image showing 44P/Reinmuth 2, made up of 2$\times$80 second exposures. The comet, in the centre of the frame, is very faint but appears to be inactive.}
         \label{44Pimage}
   \end{figure}


44P/Reinmuth 2 was expected to have $m_R$ = 22.8 at the time of our observations, and was therefore a secondary target. Only one frame was taken of this comet on our first night of observations before high humidity forced us to close the dome, making it impossible to distinguish the comet from faint background stars. We acquired two more frames on 44P on the second night. 

   \begin{figure}
   \centering
   \includegraphics[angle=-90,width=0.47\textwidth]{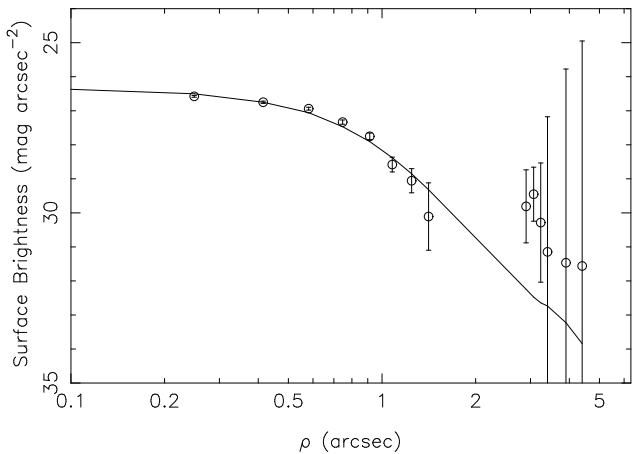}
      \caption{Surface brightness profile of 44P, showing surface brightness in magnitudes per square arcsecond against distance from the centre of the comet $\rho$ (arcsec). Beyond 2.5\arcsec{} sky noise dominates the profile, as can be seen by the large error bars, however the comet (points) and scaled stellar (solid line) profiles are indistinguishable in the inner part of the profile, implying that the comet was indeed inactive.}
         \label{44Psbp}
   \end{figure}


The comet appears stellar in each of these frames, and in a combined frame (Fig.~\ref{44Pimage}). To put limits on any unresolved coma, we measure the surface brightness $\Sigma$ at a large distance $\rho$ (arcsec) from the centre of the comet image. For a steady state coma, the surface brightness is inversely proportional to $\rho$, and the integrated coma magnitude within $\rho$, $m_c(\rho)$, is given by \citep{Jewitt+Danielson84}
\begin{equation}\label{sbp-eqn}
m_c(\rho) = \Sigma_c(\rho) - 2.5\log(2\pi\rho^2).
\end{equation}
Figure~\ref{44Psbp} shows the measured surface brightness profile, which appears stellar within the inner 2.5\arcsec, but is dominated beyond that by sky noise. We put a limit on any coma within this radius at $m_c(2.5) \ge 26.9$, corresponding to at most 2\% of the total flux from the comet. We conclude that this comet was inactive at the time of observation, which we would expect at $R_{\rm h}$ = 5.2 AU.

We measure $m_R$ = 22.62$\pm$0.12, corresponding to $r_{\rm N}$ = 1.96$\pm$0.11. This measurement is consistent with the upper limit of $r_{\rm N} \le$ 3.0 km found by \citet{Lowry03}. It is slightly larger than the value of 1.61 km quoted by \citet{Lamy-chapter}, but is in agreement if we assume that the nucleus is slightly elongated. Such a range in magnitudes would require a minimum axial ratio of $a/b$ = 1.5, which is typical of the values determined for JFC nuclei (see section~\ref{discussion}).

\subsection{Active comets}

   \begin{figure}
   \centering
   \begin{tabular}{c c}
   \includegraphics[width=0.23\textwidth]{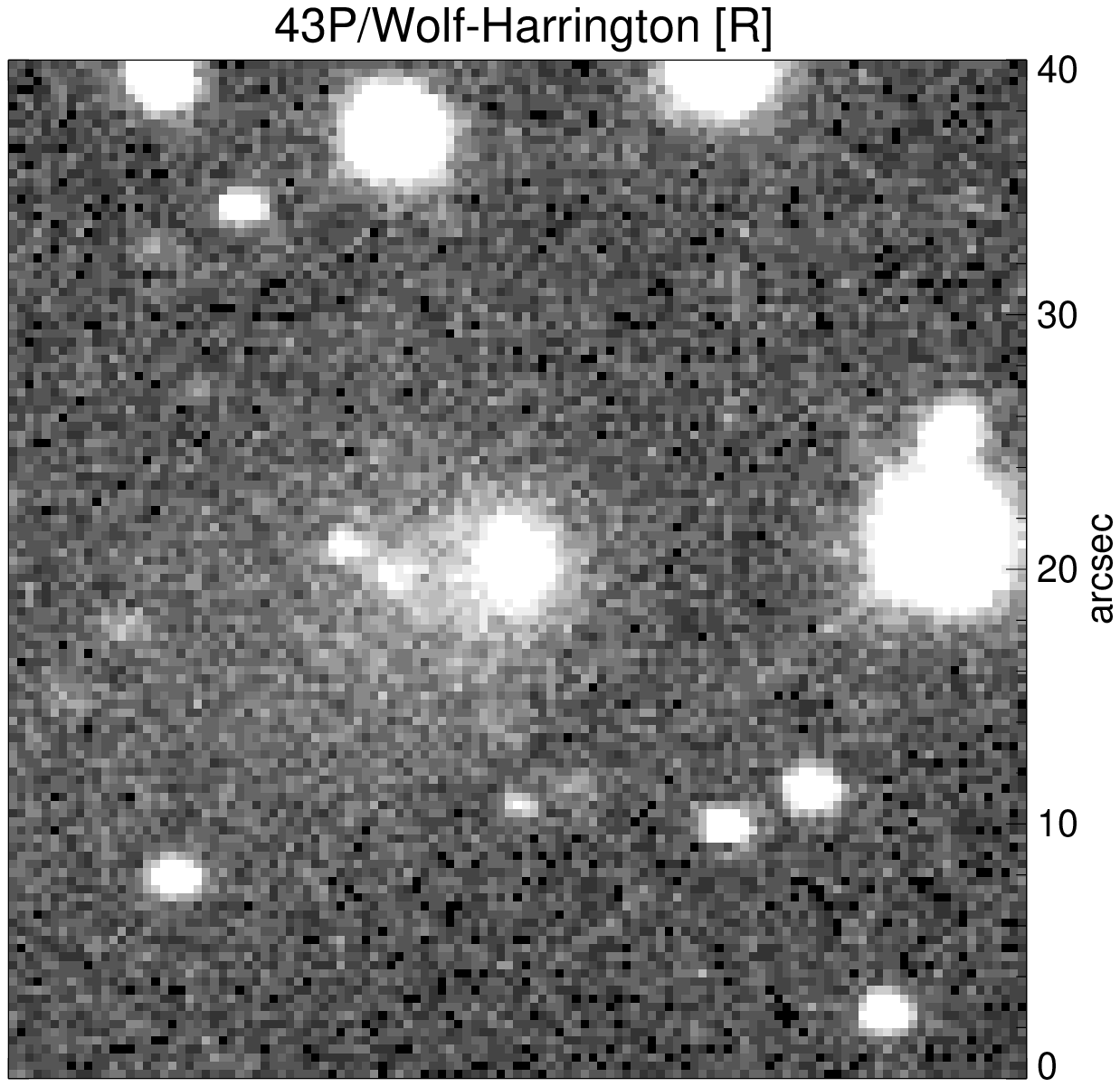} &
   \includegraphics[width=0.23\textwidth]{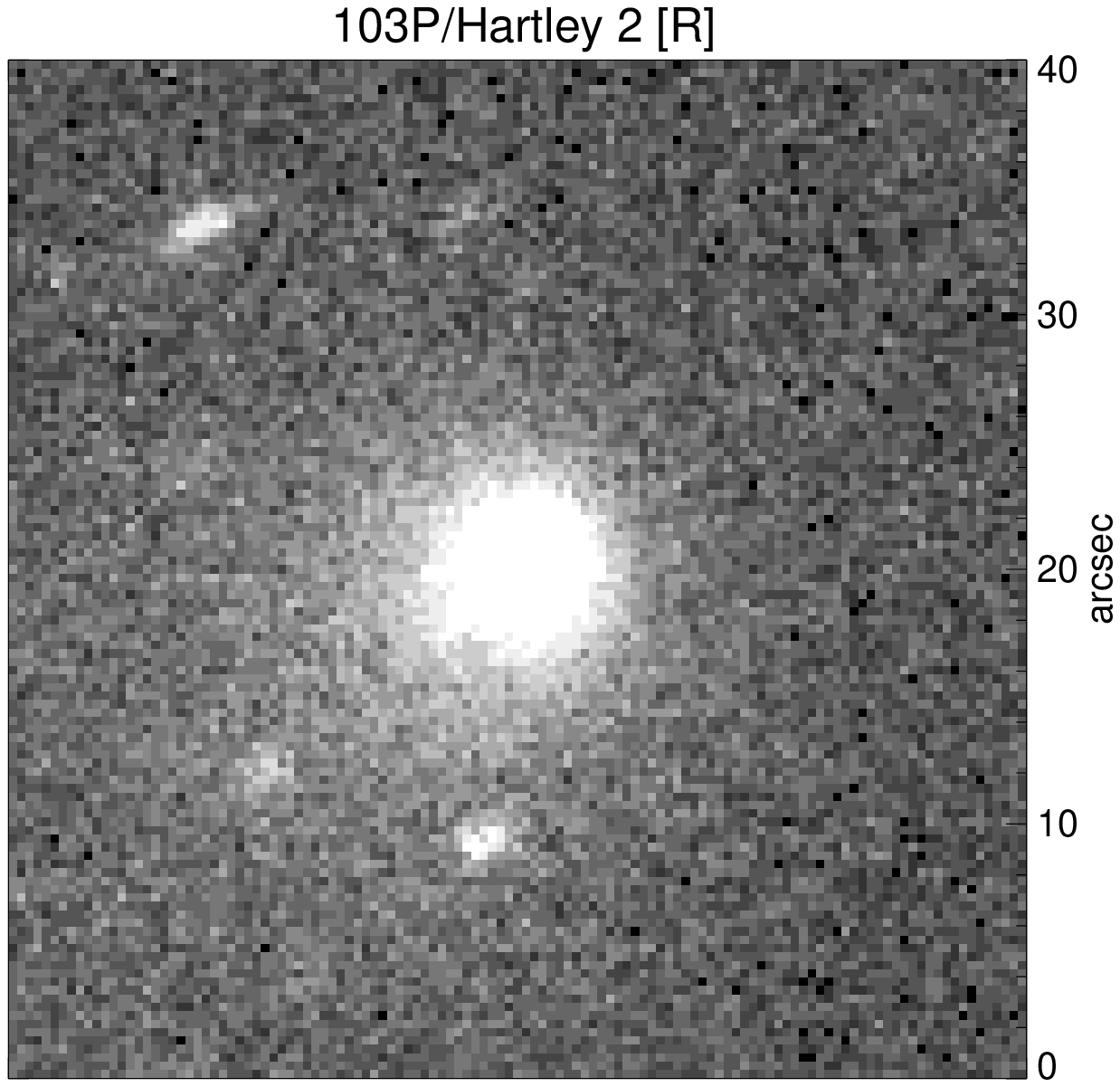} \\
   \includegraphics[angle=-90,width=0.23\textwidth]{fig3c.ps} &
   \includegraphics[angle=-90,width=0.23\textwidth]{fig3d.ps} \\
   \end{tabular}
      \caption{Images and surface brightness profiles for 43P/Wolf-Harrington and 103P/Hartley 2. The comets are visibly active. The solid diagonal lines in the top right corner of each profile show gradients of -1 and -1.5, from two theoretical models of steady state coma; both comets have profiles with gradients~$\sim -1$. Note that the error bars on the surface brightness are smaller than the data points.}
         \label{active_fig}
   \end{figure}


43P/Wolf-Harrington and 103P/Hartley 2 were both visibly active in snap-shot frames (fig.~\ref{active_fig}), and were immediately rejected as unsuitable targets for our programme of nucleus photometry. We can still place upper limits on the size of the nucleus of these comets, as the flux from the nucleus cannot be larger than the total flux measured. For 43P, we measured a total brightness within an aperture of radius 5\arcsec{} of $m_R$ = 20.53$\pm$0.03, which gives an upper limit on the effective radius of $r_{\rm N} \le$ 2.42 km. 

The level of activity of a comet can be quantified using the quantity $Af\rho$ \citep{Ahearn84}, which is roughly proportional to the dust production rate of the comet, assuming steady state production. It is the product of the geometric albedo $A$, the radius of the photometric aperture at the comet $\rho$, and the filling factor $f$, which is the total scattering cross-section of the grains within this aperture. We calculate this quantity by
\begin{equation}
Af\rho [ {\rm cm} ] = \frac{(2\Delta R_{\rm h})^2 F_{\rm comet}}{\rho F_\odot},
\end{equation}
where $R_{\rm h}$ is in AU, but $\Delta$ and $\rho$ are in cm. $F_{\rm comet}$ and $F_\odot$ are the fluxes measured at the Earth from the comet are from the Sun respectively. Expressed in this fashion, $Af\rho$ is theoretically independent of the chosen aperture. We take an aperture of radius 5\arcsec, which corresponds to $\rho$ = 9400 km for 43P, and gives $Af\rho$ = 206 $\pm$ 2 cm. This is a large value, which is expected for a comet which shows high activity levels well beyond the canonical 3 AU `cut-off' for H$_2$O sublimation. This comet's perihelion distance jumped from $q$ = 2.5 to 1.5 during 1936; it is likely that this major change in its orbit is connected to the high activity level observed.

This activity level has frustrated previous efforts to obtain a definitive measurement of the radius of the nucleus; although there have been a number of published snap-shot observations of 43P, it has mostly been observed when active, even at distances beyond 4 AU (table~\ref{43Pall}). The upper limits obtained when the comet was active are all \emph{below} the radii determined by \citet{Lowry99,Lowry03} from images in which the comet appeared inactive. There are two possible explanations for this discrepancy; firstly, that the range in reported magnitudes is due to elongation of the comet, in which case the full range of $\Delta m$ = 0.97 mag.~from the peer-reviewed literature \citep{Lowry99,Licandro00b,Lowry03,Lowry+Fitzsimmons05} gives a minimum axial ratio of $a/b \ge 10^{0.4\Delta m} = 2.4$. 
The alternative explanation is that the apparently stellar observations contained significant unresolved coma. This would be quite possible, as both were obtained with the 1.0m JKT, and the surface brightness profile and equation~\ref{sbp-eqn} reveal that $m_c \approx m_R$ in both cases, suggesting that unresolved coma could have contributed over 90\% of the flux in each of these observations.

\begin{table*}
\begin{minipage}[t]{2\columnwidth}

\caption{Reported observations of 43P/Wolf-Harrington}             
\label{43Pall}      
\centering                         
\begin{tabular}{@{}lccccccccl@{}}        
\hline                 
Date & $R_\mathrm{h}$\footnote{Superscripts $I$ and $O$ refer to whether the comet is inbound (pre-perihelion) or outbound (post-perihelion).} & $\Delta$ & $\alpha$ & $m_R$ & $m_c$ & $m_R(1,1,0)$\footnote{Recalculated here using $m_\odot$=-27.09, $A_R$=0.04 and $\beta$=0.035, for consistency.\label{fn4}} & $r_{\mathrm{N}}$\footref{fn4} & App.\footnote{Appearance is given as either A -- active, or S -- stellar.} & Ref.  \\   
& [AU] & [AU] & [deg.] & & & & [km] & & \\ 
\hline
04/02/1992 & 3.04$^O$ & 2.69 & 18.6 & 20.45 $\pm$ 0.13\footnote{From reported $m_V=20.6\pm0.1$, using $(V-R)=0.15\pm0.08$ from Lowry \& Fitzsimmons 2005} & - & $\ge$15.24 & $\le$2.57 & A & Licandro et al.~2000\\
24/08/1995 & 4.87$^I$ & 4.26 & 10.1 & 21.40 $\pm$ 0.50 & $\ge$21.48 & 14.46 & 3.67 & S & Lowry et al.~1999\\	
June 1996 & 3.9$^I$ & - & - & - & - & $\ge$16.25\footnote{From reported $m_V(1,1,0) \ge 16.4$, using $(V-R)=0.15\pm0.08$ from Lowry \& Fitzsimmons 2005} & $\le$1.61 & A & Hainaut et al.~1996\\
13/06/1999 & 4.46$^O$ & 3.66 & 9.0  & 20.81 $\pm$ 0.14 & $\ge$20.35 & 14.43 & 3.72 & S & Lowry et al.~2003\\	
12/07/2002 & 4.43$^I$ & 3.45 & 3.8  & 21.45 $\pm$ 0.06 & - & $\ge$15.40 & $\le$2.38 & A & Lowry \& Fitzsimmons 2005\\
05/03/2005 & 3.30$^O$ & 2.61 & 13.9 & 20.53 $\pm$ 0.03 & - & $\ge$15.37 & $\le$2.42 & A & this work\\

\hline                                   
\end{tabular}
\end{minipage}
\end{table*}

\citet{Lowry+Fitzsimmons05} argue for the first scenario, as Lowry's images were taken at larger $R_{\rm h}$, the two values of $r_{\rm N}$ obtained are consistent with each other and $a/b \ge 2.4$ falls within the observed range for nuclei (see section~\ref{discussion}). However, our additional limit below 2.5 km pushes the weight of probability towards the second explanation. Furthermore, if we include Hainaut et al.'s 1996 result of $m_V(1,1,0) \ge 16.4$ (MPC~27955, quoted in \citealt{Licandro00b}) and use $(V-R)=0.15 \pm 0.08$ \citep{Lowry+Fitzsimmons05} to convert to the $R$-band, the first scenario requires $\Delta m$ = 1.82, and therefore $a/b \ge 5$, which we regard as unrealistically large. Even taking the 3$\sigma$ upper limit on the colour of $(V-R)$ = 0.39 gives $r_{\rm N} \le 1.8$ km and $a/b \ge 4$. We therefore conclude that there was significant flux from unresolved coma present in Lowry et al.'s observations, and that 43P definitely has $r_{\rm N} \le 2.4$ km, and quite possibly $r_{\rm N} \le 1.6$ km.

For 103P we found $m_R$ = 18.72$\pm$0.02, implying $r_{\rm N} \le$ 4.13 km. Within $\rho$ = 8200 km we measure $Af\rho$ = 196 $\pm$ 1 cm. Such high activity is unsurprising as the comet was outbound at only 3.2 AU, and it has been observed active at 4.6 AU \citep{Lowry03}. The radius limit is in agreement with other limits of $\le 5.3$ km \citep{Licandro00b}, $\le 5.9$ km \citep{Lowry+Fitzsimmons01} and $\le 5.8$ km \citep{Lowry03}, although it has been reported that the radius is much smaller than these limits; \citet{Groussin04} give a radius of 0.71 $\pm$ 0.13 km from Infrared Space Observatory observations of the comet when it was active at $R_{\rm h}$ = 1.2 AU in February 1998.

\section{Time-series photometry}
\label{lc-results}

\subsection{17P/Holmes}

   \begin{figure}
   \centering
   \includegraphics[width=0.47\textwidth]{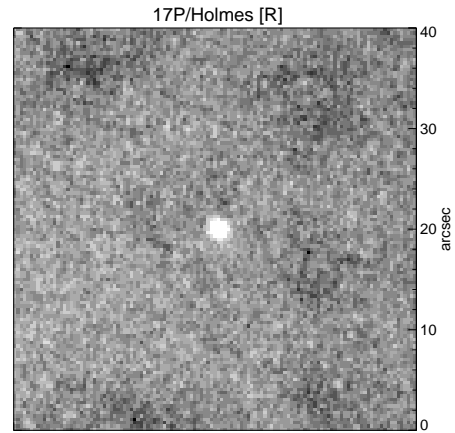}
      \caption{Image showing 17P/Holmes, made up of 25$\times$100 second exposures taken on the second night. The comet appears to be inactive. The frames are combined in such a way as to remove background sky, fixed objects and cosmic rays.}
         \label{17Pimage}
   \end{figure}

\begin{figure}
   \centering
   \includegraphics[angle=-90,width=0.47\textwidth]{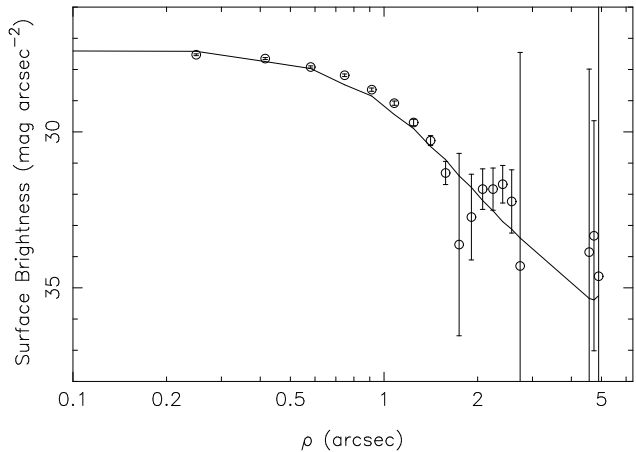}
      \caption{Same as fig.~\ref{44Psbp} for 17P. The stellar profile matches the comet, implying that the comet was inactive.}
         \label{17Psbp}
   \end{figure}

17P/Holmes appears to have been inactive at the time of observation. Figure~\ref{17Pimage} shows a combined image showing only the comet, produced by removing a scaled version of the stellar background from each frame prior to combining them, and is made up of all 25 $R$-band exposures taken on the second night, giving an effective exposure of 2500 seconds $\equiv 0.7$ hours. A surface brightness profile (fig.~\ref{17Psbp}) shows no sign of activity, and we use equation~\ref{sbp-eqn} to put a limit on any unresolved coma at $m_c \ge 24.6$, or $\le$ 13.9 $\pm$ 28.6 \% of the total flux.

We use the Fourier method described in \citet{Snodgrass05} to search for periodicities in the brightness variations of the comet (data given in table~\ref{17Ptable}). We produce a reduced $\chi^2$ periodogram, where minima with reduced $\chi^2$ in the range $\chi^2/\nu = 1 \pm \sqrt{2/\nu}$ give residuals with a Gaussian distribution with $1\sigma$ variance. $\nu$ is the number of degrees of freedom of the model, given by $\nu = (N - 3)$ where $N$ is the number of data points, for a first order Fourier model.

   \begin{figure}
   \centering
   \includegraphics[angle=-90,width=0.47\textwidth]{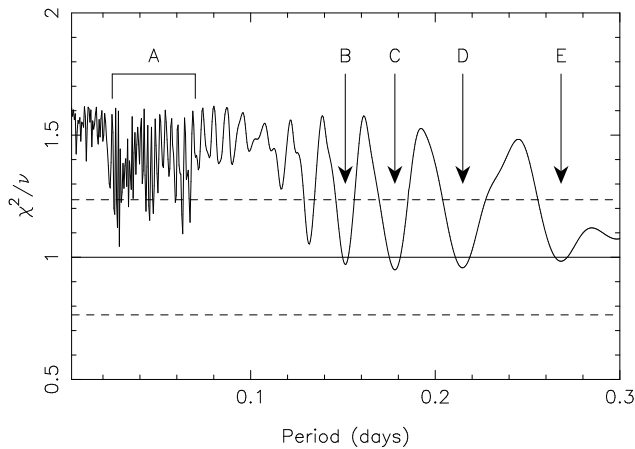}
      \caption{Periodogram for 17P. Reduced $\chi^2$ is shown against period in days. A good fit should fall within $\chi^2/\nu = 1 \pm \sqrt{2/\nu}$; the solid horizontal line is at $\chi^2/\nu = 1$, the dotted line marks the 1$\sigma$ level. There are many statistically acceptable minima: those marked `A' are at periods between 30 and 90 minutes and are due to the data sampling frequency. The data folded onto the periods with each of the four strongest minima are shown in fig.~\ref{17Pmulti}.}
         \label{17Ppgram}
   \end{figure}

\begin{figure}
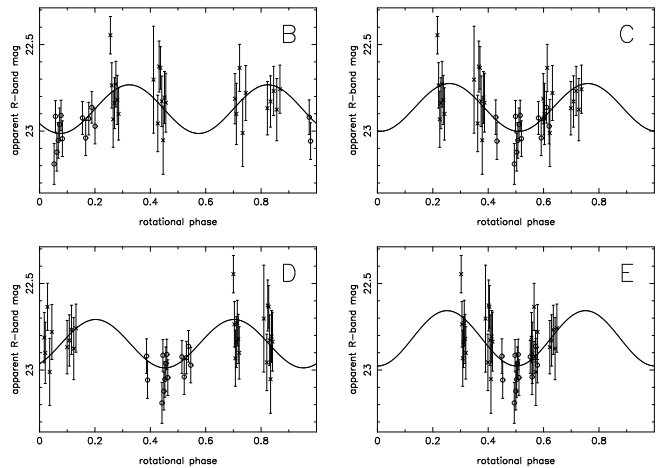

   \centering
   \begin{tabular}{c c}
   \includegraphics[angle=-90,width=0.23\textwidth]{fig7a.ps} &
   \includegraphics[angle=-90,width=0.23\textwidth]{fig7b.ps} \\
   \includegraphics[angle=-90,width=0.23\textwidth]{fig7c.ps} &
   \includegraphics[angle=-90,width=0.23\textwidth]{fig7d.ps} \\
   \end{tabular}
      \caption{17P photometric data folded onto each of the four strongest periods found in the periodogram (fig.~\ref{17Ppgram}). These have $P_{\rm rot}$ = 7.2 (B), 8.6 (C), 10.3 (D) and 12.8 (E) hours. The 1.5 nights of data acquired on this comet are insufficient to choose one of these periods over the others.}
         \label{17Pmulti}
   \end{figure}

Figure~\ref{17Ppgram} shows the periodogram for 17P, with $\chi^2/\nu$ against period in days. We have a total of 39 data points in the $R$-band, so $\sqrt{2/\nu} = \sqrt{2/(39 - 3)} = 0.236$. There are clearly a number of solutions within the dotted lines marking $\chi^2/\nu = 1 \pm 0.236$; in fact our null hypothesis (that there is no periodic variation, and all scatter is due to noise around the mean magnitude) is only rejected at the 2$\sigma$ level. With only 1.5 nights of data it is impossible to choose between the possible periods. We can reject those marked `A' in fig.~\ref{17Ppgram} as they are at periods between half and 1.5 hours, and are due to the various aliases created by observing the comet in blocks with temporal separation of that order. In fig.~\ref{17Pmulti} we show the data folded onto the rotation periods corresponding to each of the four periods with $\chi^2/\nu \approx 1$ (marked B--E), and note that they each produce a visually acceptable light-curve. 

We assume that the variation in brightness is due to the nucleus being a rotating non-spherical body, and therefore expect to see a double peaked light-curve with a rotation period $P_{\mathrm{rot}}$ twice the fitted Fourier period $P_{\mathrm{fitted}}$. The four minima are found at $P_{\mathrm{fitted}}$ = 3.6 (B), 4.3 (C), 5.2 (D) and 6.4 (E) hours, and correspond to rotation periods of $P_{\mathrm{rot}}$ = 7.2, 8.6, 10.3 and 12.8 hours. 

We plot the data from night 1 as open circles and the night 2 data as crosses. While the period remains uncertain, it can be seen that we observed 17P at a light-curve minimum on the first night, and at around maxima on the second. Unfortunately there is a gap in the data on the second night, due to the comet passing in front of a star, which appears to have coincided with the minimum between these maxima, assuming that one of the above periods is correct. We are confident that the calibration of the data from the two different nights is accurate, as repeating the method with different comparison stars produced no significant difference in the calibrated magnitudes; the uncertainty on the night-to-night calibration is much smaller than the error bars on individual points. Therefore we can be confident that we have observed the full amplitude of the light-curve at this epoch. Here we follow the common practice of describing the nucleus as the simplest non-spherical shape: a tri-axial ellipsoid with semi-axes $a \ge b = c$. The variations in brightness that produce the light-curve are then due to the changing cross-section of the rotating elongated nucleus. We can only measure a lower limit on $a/b$, as we do not know the orientation of the rotation axis. From $\Delta m$ = 0.3 we obtain a lower limit on the elongation of the nucleus of $a/b \ge 1.3$.

We measure the mean apparent magnitude of 17P to be $m_R = 22.86 \pm 0.02$. Using equation~\ref{rneqn}, this gives an effective radius of $r_{\rm N} = 1.62 \pm 0.01$ km for the equivalent spherical body. Taking $a/b = 1.3$, this corresponds to dimensions of the nucleus of $a\times b = 1.9\times 1.4$ km. This is in excellent agreement with the $r_{\rm N}$ = 1.71 km measurement from a snap-shot observation acquired by \citet{Lamy00} using the HST. 

Measuring the elongation from the range of the light-curve requires that the variations are in fact due to changes in the projected surface area of the comet, and not due to large scale changes in albedo. To test this assumption, we measured the colour indices $(V-R)$ and $(R-I)$ at least once per night, to check for any colour variations which would indicate varying surface properties. We acquired `colour blocks' by cycling through the three filters $VR$ and $I$ in the following order: $RVRIR$, repeating the last four frames for faint objects. This allowed direct interpolation to find the $R$-band magnitude at the time of the $V$ or $I$-band observations, by simply averaging the $R$-band frames bracketing them. For 17P we acquired one colour block, containing 2$\times$100s $V$ and $I$-band frames in the sequence $RVRIRVRIR$, on the first night and two blocks on the second. The colours measured were consistent within the uncertainties on each, supporting our assumption that variations are due to the shape of the nucleus, and give average colours of $(V-R) = 0.41 \pm 0.07$ and $(R-I) = 0.44 \pm 0.08$.

\subsection{47P/Ashbrook-Jackson}

   \begin{figure}
   \centering
   \includegraphics[width=0.47\textwidth]{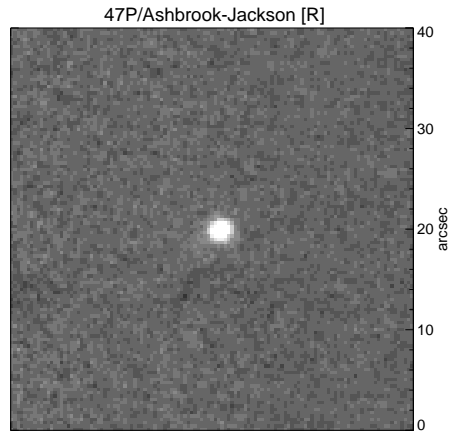}
      \caption{Image showing 47P/Ashbrook-Jackson, made up of 34$\times$85 second exposures taken on the second night. The comet appears to be inactive.}
         \label{47Pimage}
   \end{figure}

\begin{figure}
   \centering
   \includegraphics[angle=-90,width=0.47\textwidth]{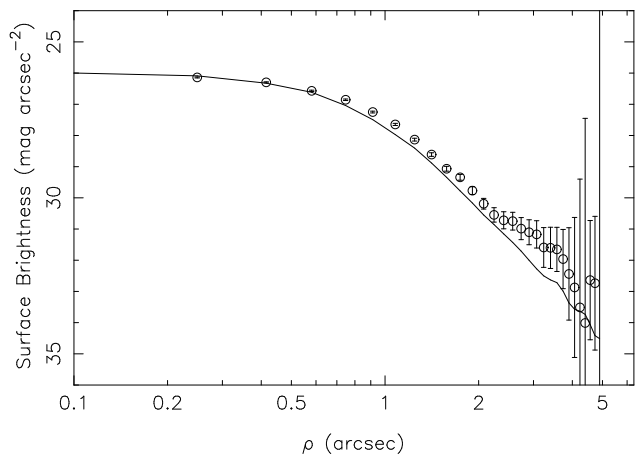}
      \caption{Same as fig.~\ref{44Psbp} for 47P. The stellar profile matches the comet, implying that the comet was inactive.}
         \label{47Psbp}
   \end{figure}

47P/Ashbrook-Jackson was visible throughout the entire time we were able to observe on both nights, and consequently was well monitored, with a total of 53 $R$-band exposures of 85 seconds each (table~\ref{47Ptable}). The comet appeared bright and stellar in each; a combination of all of the frames taken on the second night is shown in fig.~\ref{47Pimage}. A surface brightness profile (fig.~\ref{47Psbp}) shows that the comet has negligible coma, with a scaled stellar profile matching the comet down to surface brightness $\Sigma > 30$ mag arcsec$^{-2}$. Using equation~\ref{sbp-eqn} we put limits on any unresolved coma at $m_c \ge 27.4$, implying that activity is entirely negligible and can contribute no more than $0.5 \pm 2.1$\% of the total flux, as was expected as this comet was close to aphelion at the time of observation.

   \begin{figure}
   \centering
   \includegraphics[angle=-90,width=0.47\textwidth]{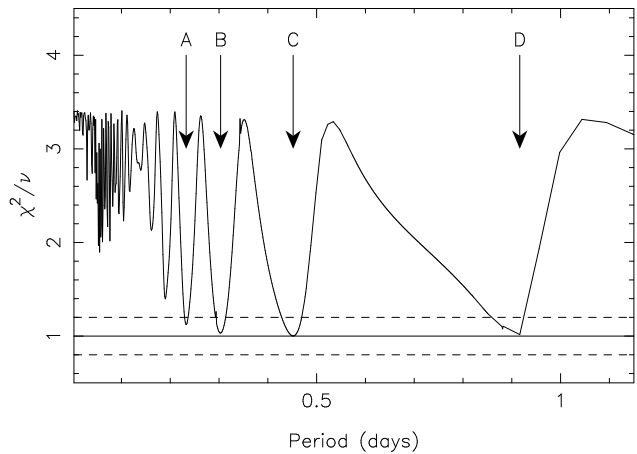}
      \caption{Periodogram for 47P. There are four statistically acceptable minima: the data folded onto the periods with each of these minima are shown in fig.~\ref{47Pmulti}.}
         \label{47Ppgram}
   \end{figure}

\begin{figure}
   \centering
   \begin{tabular}{c c}
   \includegraphics[angle=-90,width=0.23\textwidth]{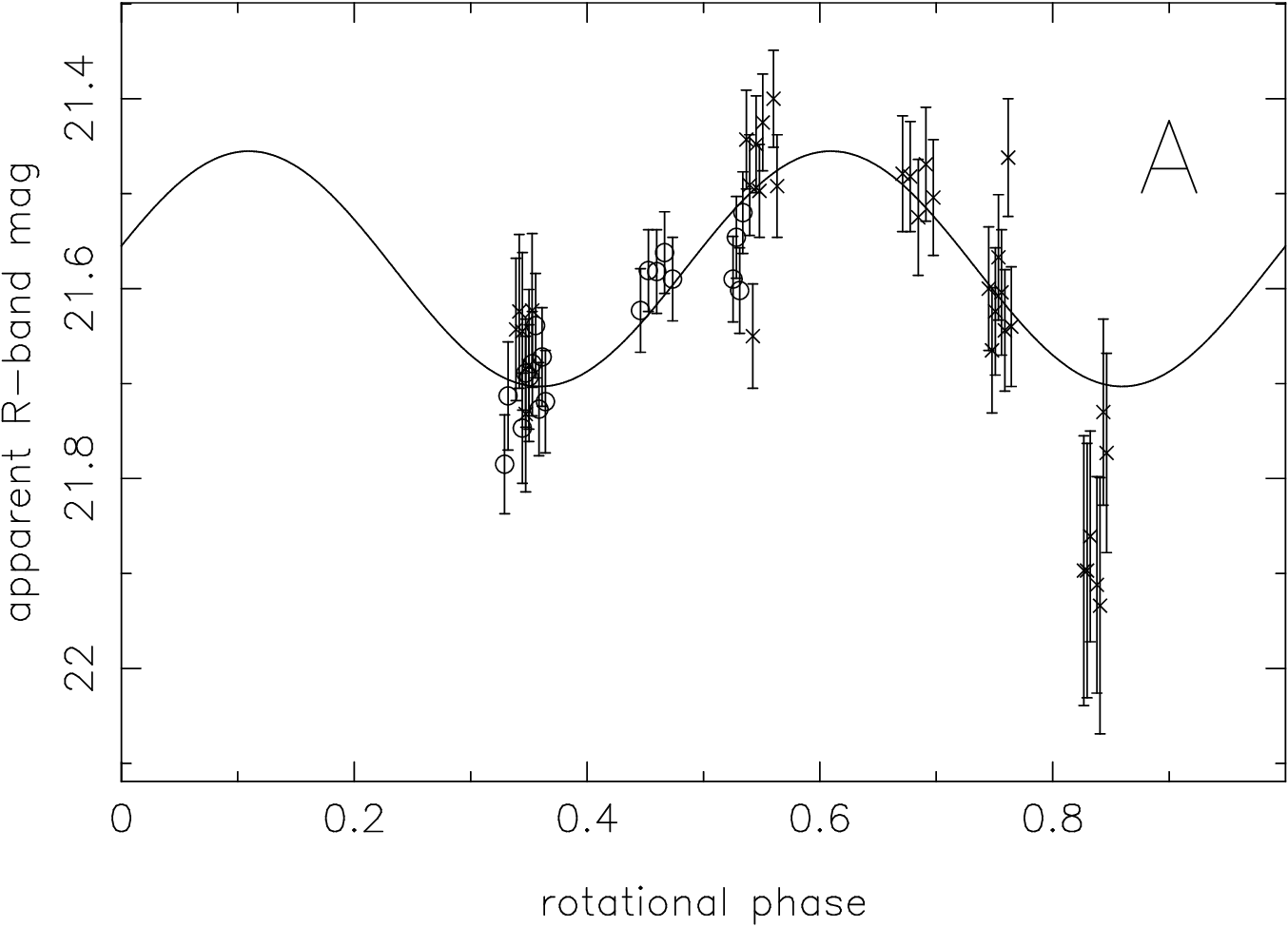} &
   \includegraphics[angle=-90,width=0.23\textwidth]{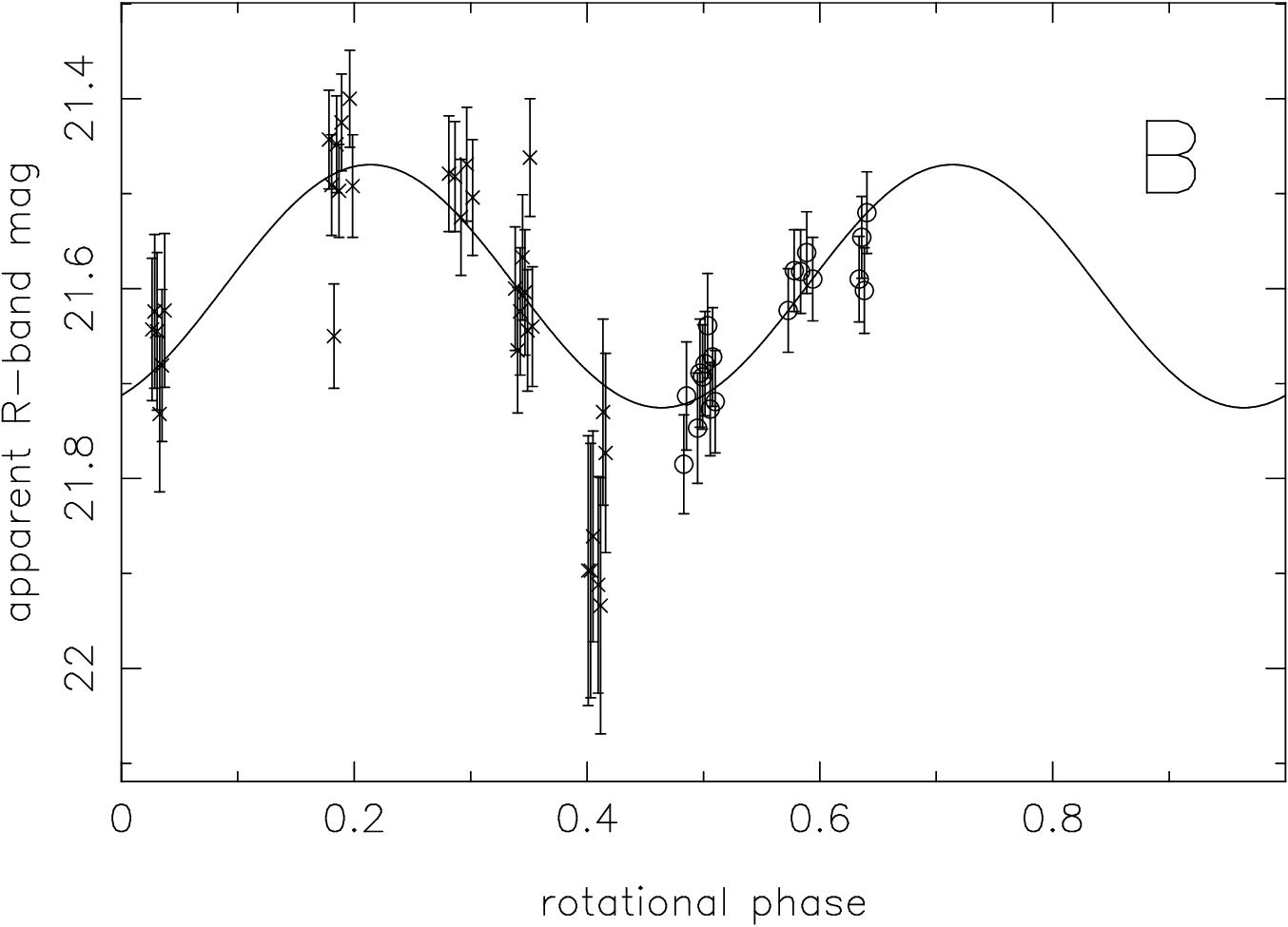} \\
   \includegraphics[angle=-90,width=0.23\textwidth]{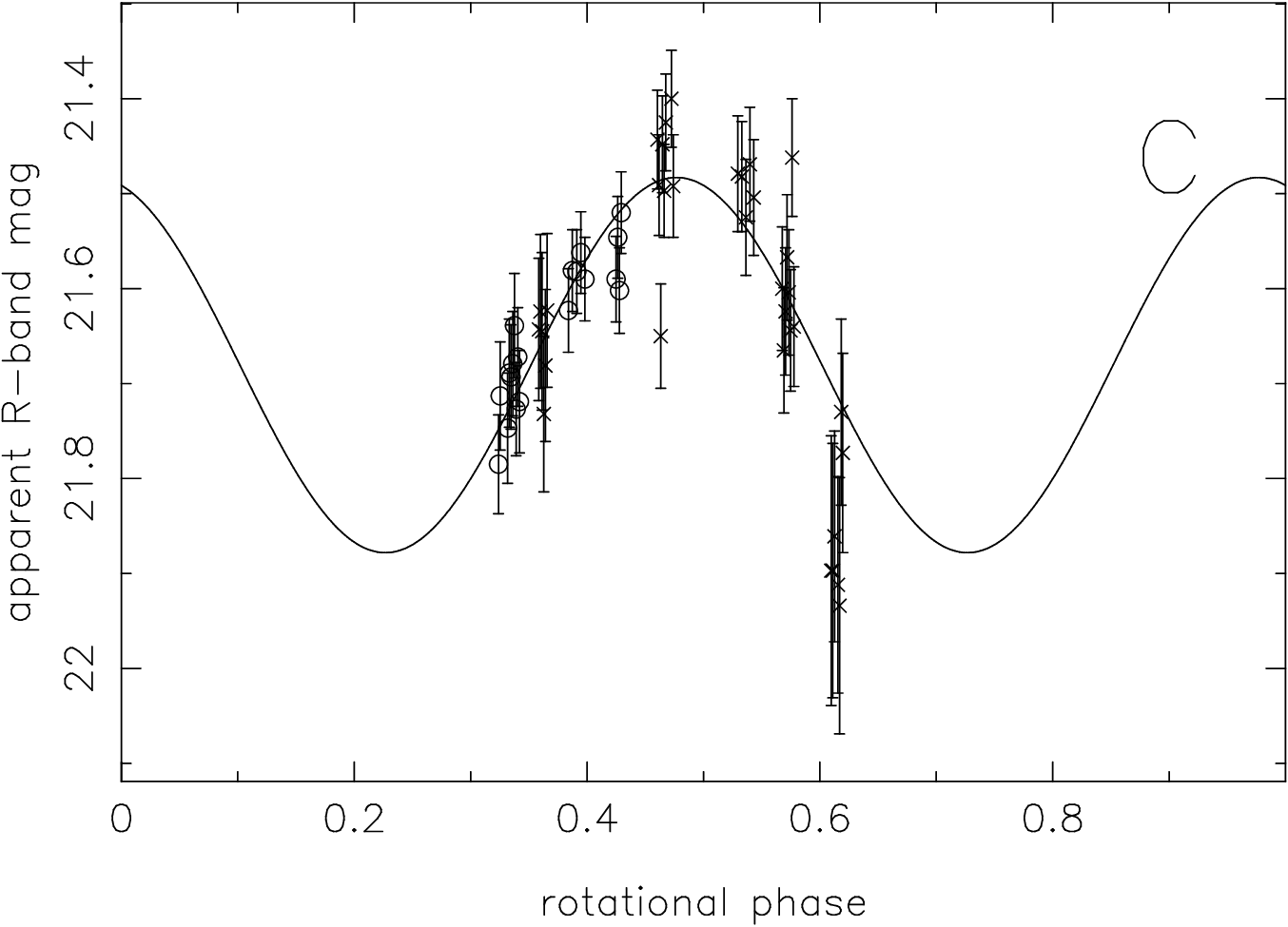} &
   \includegraphics[angle=-90,width=0.23\textwidth]{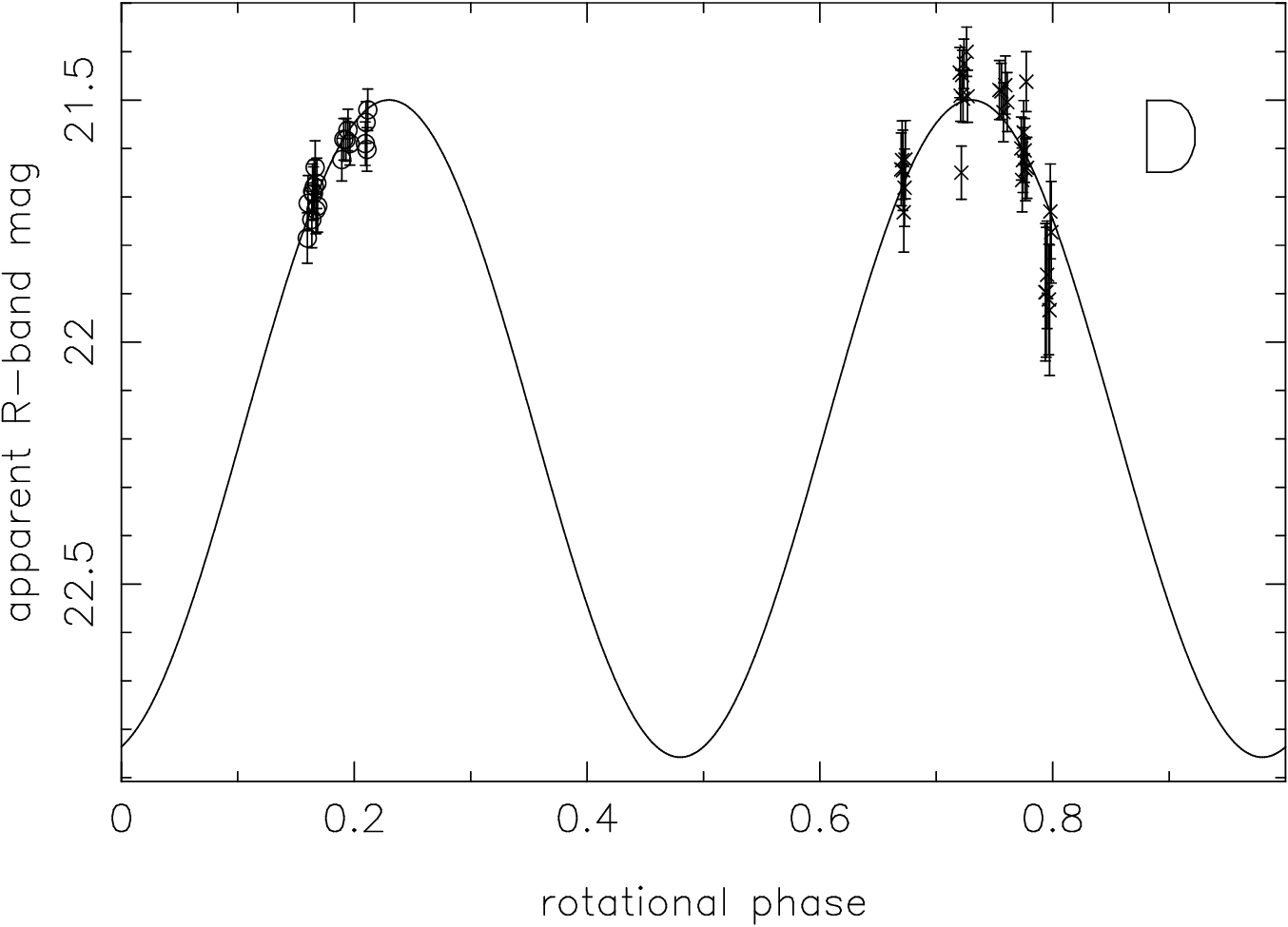} \\
   \end{tabular}
      \caption{47P photometric data folded onto each of the four strongest periods found in the periodogram (fig.~\ref{47Ppgram}). These have $P_{\rm rot}$ = 11.2 (A), 15.5 (B), 21.6 (C) and 44.0 (D) hours. The 1.5 nights of data acquired on this comet are insufficient to choose one of these periods over the others.}
         \label{47Pmulti}
   \end{figure}

There are four statistically acceptable minima in the periodogram for 47P, which are marked A--D in fig.~\ref{47Ppgram}. Light-curves showing the data folded onto each of these are shown in fig.~\ref{47Pmulti}. Period `C' ($P_{\rm rot} = 21.6$ hrs) is marginally the strongest minima, but the data do not allow us to choose between these periods. In this figure it can clearly be seen that over the course of the second night a definite maximum was observed, and the approach to a maximum appears to have been observed on the first night, which puts strong constraints on the rotation period, although we cannot choose between the four possible periods shown based on these data. Note that the data do allow us to measure the period precisely, with small error bars on the determined periods of $11.2\pm0.3, 15.5\pm0.5, 21.6\pm1.0$ and $44.0^{+1.0}_{-2.9}$ hours, but do not give a unique solution. It is clear that the periods are aliases of each other at $\sim 1, 3/2, 2$ and $4 \times 11$ hours, yet it is not possible to select one as the correct period. The rotation period cannot be longer than 44 hours (period `D'), as this period requires that we saw alternate maxima of a double peaked light-curve on the 2 nights; the remaining solutions imply that there were other maxima unobserved between these events. There is an unpublished partial light-curve on 47P acquired by Lamy et al. using the HST, which puts a lower limit on the rotation period at 44.5 hours. If this is correct, it would suggest that period `D' is the real one, and that the shape of the body is irregular: The `best-fit' sinusoid drawn through the data to guide the eye in fig.~\ref{47Pmulti}(D) does not represent the data well. The range in magnitudes from the data is $\Delta m$ = 0.45, giving $a/b \ge 1.5$, which is typical of JFC nuclei.

The mean magnitude was measured to be $m_R = 21.68 \pm 0.01$, which gives a radius of 3.38 $\pm$ 0.01 km, assuming an albedo of 4\%. Taken with the elongation implied by the full range in the data, $a/b \ge 1.5$, we get dimensions of the nucleus of $a\times b = 4.2\times2.8$ km, for what is most likely an overly simple model of the nucleus. This is in agreement with the radius measurements of 2.8 km \citep{Lamy-chapter}, 3.1 km \citep{Licandro00b} and the limit of $r_{\rm N} \le 6.1$ km found by \citet{Lowry03}. The colours measured were $(V-R) = 0.45 \pm 0.03$ and $(R-I) = 0.38 \pm 0.03$.

\subsection{137P/Shoemaker-Levy 2}

   \begin{figure}
   \centering
   \includegraphics[width=0.47\textwidth]{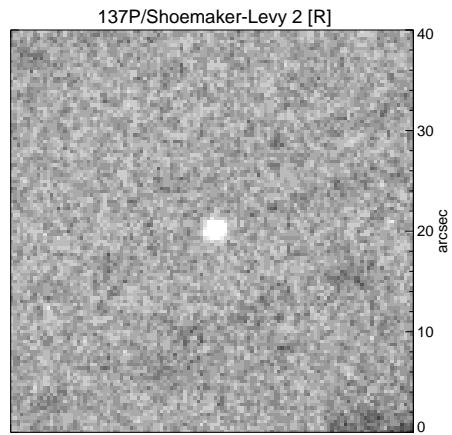}
      \caption{Image showing 137P/Shoemaker-Levy 2, created from 22$\times$140 second exposures taken on the second night.}
         \label{137Pimage}
   \end{figure}

\begin{figure}
   \centering
   \includegraphics[angle=-90,width=0.47\textwidth]{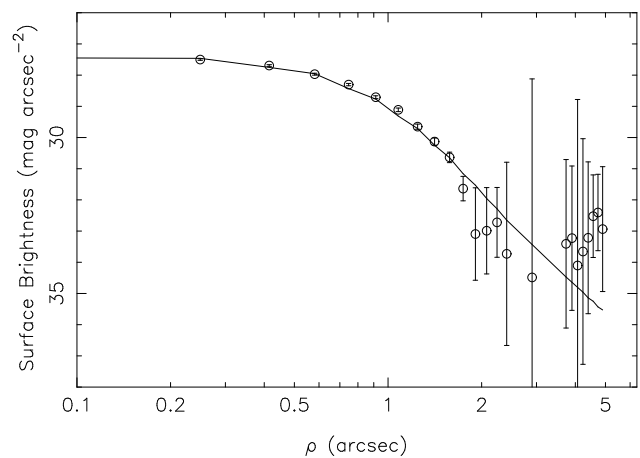}
      \caption{Same as fig.~\ref{44Psbp} for 137P. The stellar profile matches the comet, implying that the comet was inactive.}
         \label{137Psbp}
   \end{figure}

137P/Shoemaker-Levy 2 was a secondary target, as its extremely large distance (7.0 AU, inbound) gave it a predicted brightness of $m_R = 23.0$, making it a challenging target even with the 3.6m NTT. However, we attempted a snap-shot on the second night of observations at a point when all our primary targets were either below the horizon or too close to field stars. The comet was detected at a similar brightness to 17P, so was added to our light-curve target list. An image produced by shifting and adding all $R$-band data and the corresponding profile (figs.~\ref{137Pimage} \&~\ref{137Psbp}) show that the comet was inactive, as expected as it was close to aphelion. The limit on any coma contribution within a 5\arcsec{} aperture is formally found to be $m_c \ge 22.9$, which corresponds to up to 82\% of the flux, however at $\rho = 5\arcsec$ sky noise entirely dominates; although most of the background sky is removed by the process that gives fig.~\ref{137Pimage}, the comet profile can not be measured accurately beyond $\Sigma = 32$ mag arcsec$^{-2}$. The profile is clearly stellar in the inner part; measuring the surface brightness at $\rho = 2\arcsec$ gives $m_c \ge 25.0$, corresponding to $\le 14 \pm$ 10\% of the flux.

   \begin{figure}
   \centering
   \includegraphics[angle=-90,width=0.47\textwidth]{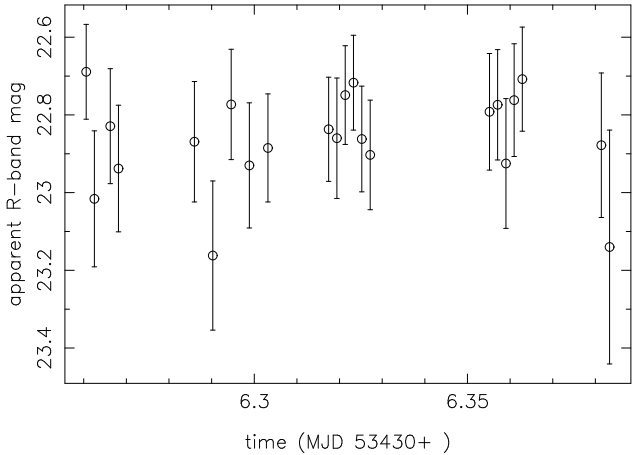}
      \caption{Photometric data for 137P. There is no statistically significant variation within the data; in the one night of data we find no evidence for rotation.}
         \label{137Porig}
   \end{figure}

We took 22 $R$-band images of 137P, however the loss of the 3rd night to bad weather meant that this data covered only 3 hours (table~\ref{137Ptable}). There is no statistically significant variation within this data, therefore we cannot constrain the rotation period. We show the data in fig.~\ref{137Porig}. This time-series allows us to obtain an accurate measurement of the brightness of this inactive nucleus, and therefore a size, again assuming a typical 4\% albedo. We measure $m_R = 22.86 \pm 0.03$, and therefore $r_{\rm N} = 3.58 \pm 0.05$ km, which is similar to the limit of $r_{\rm N} \le 3.4$ km measured by \citet{Lowry03}. We also measured colour indices for this nucleus of $(V-R) = 0.71 \pm 0.18$ and $(R-I) = 0.54 \pm 0.15$, which places 137P at the red end of the distribution of measured nuclei colours, admittedly with relatively large error bars due to its faintness.

\setcounter{table}{6}

\begin{table*}
\begin{minipage}[t]{2\columnwidth}

\caption{Derived physical parameters and colours from time series photometry on three comets.}             
\label{results_lightcurve}      
\centering                         
\begin{tabular}{@{}lccccccccc@{}}        
\hline
Comet & $m_R$ & $m_c$ & $m_R(1,1,0)$ & $r_{\mathrm{N}}$ & $P_{\mathrm{rot}}$\footnote{The rotation periods of both 17P and 47P are found to be one of four discrete values within the given ranges (see text).} & $a/b$\footnote{Lower limits as the orientation of the rotation axis is unknown.\label{fn5}} & $D_{\mathrm{N}}$\footref{fn5} & $(V-R)$ & $(R-I)$\\    
& & & & [km] & [hr] & & [g cm$^{-3}$] & & \\
\hline                        

17P & 22.864$\pm$0.020 & $\ge$24.6 & 16.241$\pm$0.020 & 1.62$\pm$0.01 & 7.2---12.8 & 1.3$\pm$0.1 & 0.09$\pm$0.02 & 0.41$\pm$0.07 & 0.44$\pm$0.08\\

47P & 21.680$\pm$0.007 & $\ge$27.4 & 14.638$\pm$0.007 & 3.38$\pm$0.01 & 11.2---44.0 & 1.5$\pm$0.1 & 0.01$\pm$0.01 & 0.45$\pm$0.03 & 0.38$\pm$0.03\\

137P & 22.862$\pm$0.032 & $\ge$22.9 & 14.513$\pm$0.032 & 3.58$\pm$0.05 & n/a & n/a & n/a & 0.71$\pm$0.18 & 0.54$\pm$0.15\\

\hline                                   
\end{tabular}
\end{minipage}
\end{table*}

\section{Discussion of comet properties and comparison with KBOs}\label{discussion}

\subsection{Bulk densities}

   \begin{figure}
   \centering
   \includegraphics[angle=-90,width=0.47\textwidth]{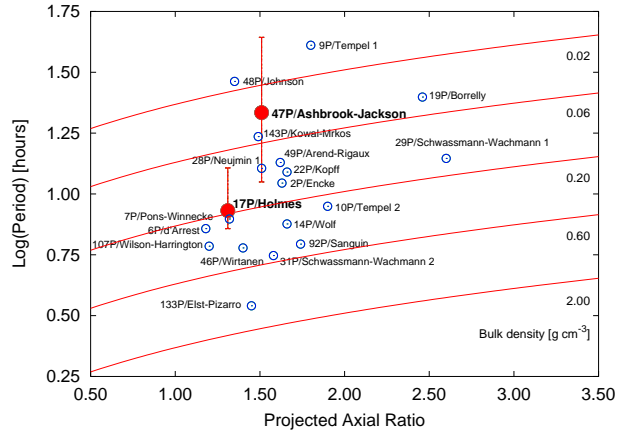}
      \caption{Rotation period against projected axial ratio for all JFC nuclei. The error bars on 17P and 47P cover the full range of possible periods discussed in the text; the points shown are for the nominally `best' periods. The solid lines show where a strengthless body would lie on these axes for a variety of densities. There appears to be a cut-off in densities at $\sim 0.6$ g cm$^{-3}$, as suggested by \citet{Lowry+Weissman03}. Data on other comets from \citealt[ and references therein]{Snodgrass05}.
                    }
         \label{densityplot}
   \end{figure}

We can use the rotation period and elongation found for 17P and 47P to put limits on the bulk density of the nucleus $D_{\rm N}$ (g cm$^{-3}$). By assuming that nuclei have negligible tensile strength [as evidenced by the break up of D/Shoemaker-Levy 9 under the gravitational influence of Jupiter \citep{Asphaug+Benz96} and results from the {\it Deep Impact} mission to comet 9P/Temple 1 \citep{Ahearn05}; see also \citet*{Weissman-chapter}] a lower limit on the density can be found by balancing self gravity with centrifugal force. We use the approximation given by \citet{Pravec+Harris00};
\begin{equation}\label{densityeqn}
D_{\mathrm{N}} \ge \frac{10.9}{P^2_\mathrm{rot}} \frac{a}{b}
\end{equation}
where $P_{\mathrm{rot}}$ is in hours. This equation assumes that the gravitational acceleration at the end of the long axis $a$ of the ellipsoid is the same as that for a sphere, but reduced by a factor of $b/a$. It provides an acceptable approximation to the more complex solution found by integrating over the volume of an ellipsoid \citep*{Richardson05}
\begin{equation}\label{jl_eqn}
\frac{1}{P_\mathrm{rot}}\sqrt{\frac{2\pi}{G D_{\mathrm{N}}}}  = \frac{1}{e^{3/2}} \sqrt{(e^2 -1)\left( 2e + \ln{\frac{1-e}{1+e}} \right)}
\end{equation}
where $G$ is the gravitational constant and $e = \sqrt{1 - (\frac{b}{a})^2}$. The derivation of this equation is given in two different ways by \citet{Richardson05} and \citet{Harris02}; it is also entirely equivalent to the expression given by \citet{Luu+Jewitt92} [their equation 9]. 

Using equation~\ref{densityeqn} we can calculate the minimum bulk density of each nucleus using the light-curve derived parameters $a/b$ and $P_{\rm rot}$. The calculated densities, along with all other results from the time-series data, are presented in table~\ref{results_lightcurve}. The full range of possible periods give densities $D_{\rm N} \ge$ 0.09 g cm$^{-3}$ for 17P, and $D_{\rm N} \ge$ 0.01 g cm$^{-3}$ for 47P. Even if spinning at the shortest of the possible periods found for these comets, they do not require densities larger than 0.28 and 0.13 g cm$^{-3}$. All of these lower density limits are consistent with the range observed for other nuclei; in this respect the nuclei of 17P and 47P are fairly typical, for any of the possible values of their rotation periods. The relatively long periods do not put strong constraints on their individual densities, however considering all measurements obtained we can infer valuable information concerning the densities of the JFC population. Under the assumption that all nuclei have a similar $D_{\rm N}$, then the values obtained by the light-curve technique will generally be below this value, but none will be observed above it. For a large enough sample of a population with randomly inclined rotation axes and a spin rate distribution, any observed upper limit to the minimum densities derived will give the true density of the nuclei. Figure~\ref{densityplot} shows the logarithm of the rotation period against projected axial ratio for all JFC nuclei. The plot shows that there appears to be a cut-off in densities at around 0.6 g cm$^{-3}$. This agrees with the results from the {\it Deep Impact} mission, which found comet 9P/Temple 1 to have a low density of $0.4 \pm 0.3$ g cm$^{-3}$ \citep{Richardson06}. In asteroids, there is a clear cut-off for bodies with $r > 75$ m at $\sim$ 3 g cm$^{-3}$ \citep{Pravec02}, and it has been suggested \citep{Lowry+Weissman03,Weissman-chapter,Snodgrass05} that the 0.6 g cm$^{-3}$ cut-off for comets is equivalent to this, the lower value reflecting on the fact that comets are under-dense when compared to asteroids, due to their larger volatile content and/or more porous structure than asteroids. 

The relative proportions of ice and rock within a body and the porosity cannot be found independently, however we can express the fraction of a body, by mass, made up of ice in terms of $D_{\rm N}$ and the porosity $\psi$, which is the fraction of void space within the total volume:
\begin{equation}\label{porosity_eqn}
\frac{M_{\rm ice}}{M_{\rm N}} = \frac{1 - (1-\psi)D_{\rm rock}/D_{\rm N}}{1 - D_{\rm rock}/D_{\rm ice}},
\end{equation}
where $D_{\rm ice}$ and $D_{\rm rock}$ refer to the densities of the volatile ices ($\sim$ 1 g cm$^{-3}$) and silicates ($\sim$ 3 g cm$^{-3}$) respectively. Using this equation for a range of assumed values of $\psi$ allows us to calculate the corresponding proportions of ice and rock that would be required to give a density of 0.6 g cm$^{-3}$. Inserting this value for $D_{\rm N}$ into equation~\ref{porosity_eqn} gives the mass fraction of ice of $M_{\rm ice}/M_{\rm N} = 2 - 2.5\psi$, and thus a minimum value for the porosity of $\psi = 40\%$ for an entirely ice nucleus. This is considerably higher than the values found through lab-based studies of meteorites, which have an average micro-porosity around 10\%, with the most porous having $\psi \approx 30\%$ \citep{Britt+Consolmagno03}, suggesting considerable macro-porosity or voids within nuclei. A value of $\psi = 50\%$ requires that the nucleus is 75\% ice to give $D_{\rm N}$ = 0.6 g cm$^{-3}$, while a nucleus with this density and a porosity of $\psi \ge 60\%$  must be silicate dominated (less than half of its mass as ice). In the limiting case of a comet almost entirely depleted in volatiles ($M_{\rm ice} \to 0$), a $\sim 100\%$ silicate nucleus must have $\psi \approx 80\%$ to give the low observed density. 

There is one object which has a minimum density greater than 0.6 g cm$^{-3}$ in fig.~\ref{densityplot}; 133P/Elst-Pizarro. The presence of 133P does not present a strong case against the 0.6 g cm$^{-3}$ cut-off, as this object is thought to be an asteroid-comet transition object; it was first found and numbered as asteroid 7968, and has an asteroid-like orbit, but has subsequently been found to have a dust trail along its orbit, and shows sporadic activity \citep*{Hsieh04,Toth06}. This and similar objects have been shown to be from a dynamically and possibly chemically distinct population \citep{Hsieh+Jewitt06}. Their stable orbits imply that these Main Belt comets have remained at or near their formation site. Due to the observed compositional gradient in the Solar System, this would imply that they possess a significantly higher silicate content than the JFCs that formed in the Kuiper Belt, and hence a higher density. Therefore we disregard 133P when discussing the global properties of he JFC population alone.

   \begin{figure}
   \centering
   \includegraphics[angle=-90,width=0.47\textwidth]{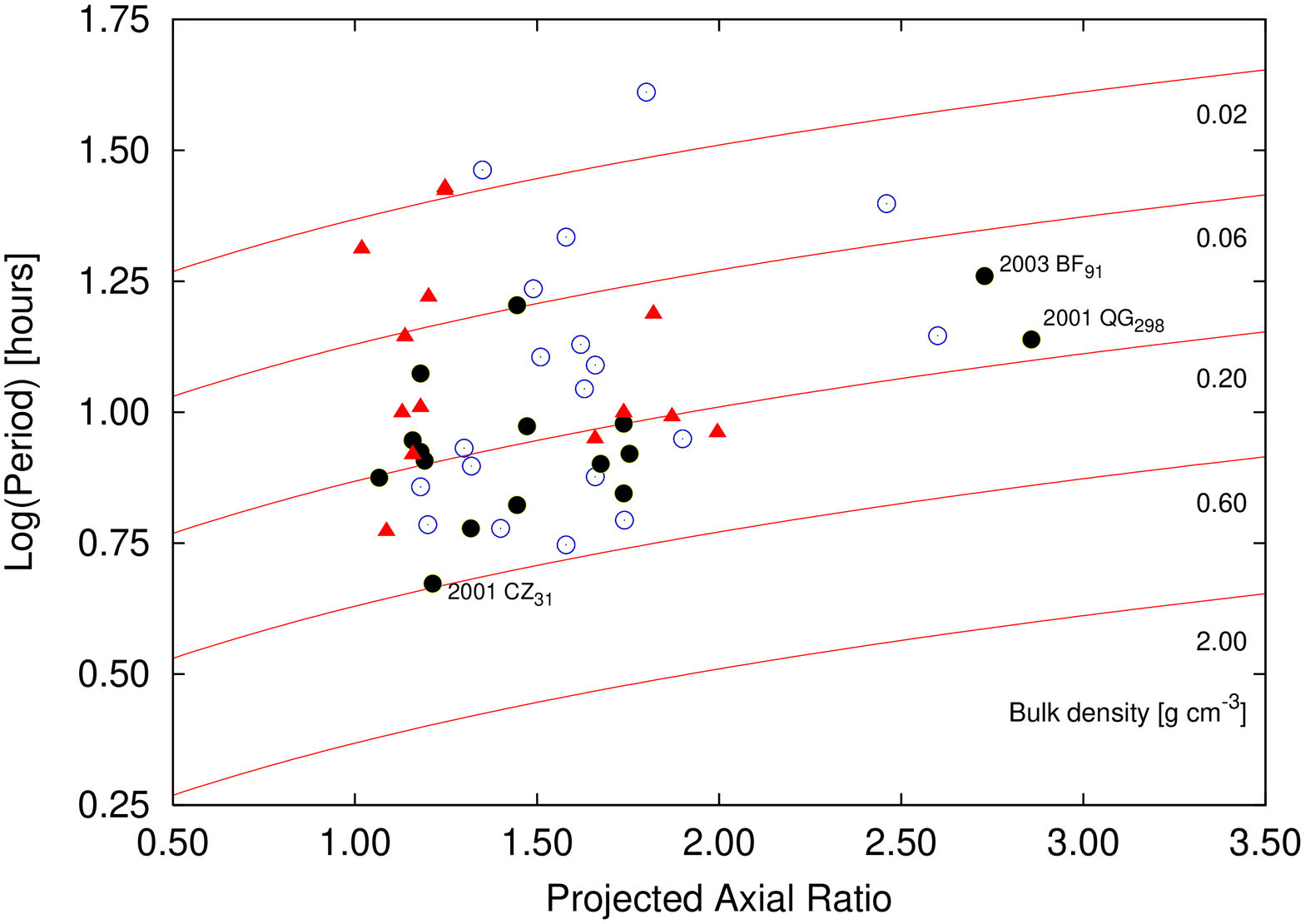}
      \caption{Same as fig.~\ref{densityplot}, but including data on KBOs (filled circles) and Centaurs (triangles) from \citet{Lacerda+Luu06}, \citet{Trilling+Bernstein06} and \citet{Ortiz06}. The distribution of JFCs is very similar to that of KBOs/Centaurs, although the KBOs tend to have lower axial ratios, most likely due to the larger size of known KBOs. The KBO data also falls above the 0.6 g cm$^{-3}$ density line, and taken together with the JFC data support the hypothesis of a cut-off at this density. The labelled objects are referred to in the text.
                    }
         \label{KBOdensityplot}
   \end{figure}

We compare the position of JFC nuclei in fig.~\ref{densityplot} with the results from similar studies on their supposed parent population, the KBOs. We select KBOs with $H_V \ge 5.1$ (corresponding to bodies with radius $\le 200$ km assuming a 10\% albedo typical of KBOs) from the recently published compilations of all KBO data by \citet{Lacerda+Luu06} and \citet{Trilling+Bernstein06} [see these papers for references to results for individual objects], and also the new results of \citet{Ortiz06}. Figure~\ref{KBOdensityplot} shows the position of the 16 KBOs (filled circles) and 14 Centaurs \& Scattered disk objects (triangles) with measured light-curves from these papers together with the JFC nuclei from fig.~\ref{densityplot}. The rotation periods plotted assume that the changes in brightness are due to the observed cross-section of a rotating elongated body (i.e. double peaked light-curves). Furthermore, to allow comparison between the nuclei and the KBOS, we must assume that the KBOs are also effectively strengthless. Whether this assumption can be applied to such large bodies is an interesting question, however the work of \citet{Davidsson99} shows that for similar rotation periods, larger bodies will fall into the `damaged region' which have been fractured due to sheer forces and are therefore only gravitationally bound. For the very largest bodies there is likely to be considerable non-homogeniety due to high central pressures. For this reason we restrict the KBOs to the smaller ones for which light-curves are available.

It is clear from fig.~\ref{KBOdensityplot} that the comets and the outer solar system bodies populate a similar region of axial-ratio--period space, which implies that they have similar bulk densities if it is assumed that neither have significant strength, supporting the theory that JFCs come from the Kuiper Belt. We note that the KBOs also have minimum bulk densities below the postulated 0.6 g cm$^{-3}$ cut-off; the fastest rotating KBO known is 2001 CZ$_{31}$, which has a rotation period of $P_{\rm rot}$ = 4.7 hours \citep{Lacerda+Luu06}, and therefore a density of $D_{\rm N} \ge 0.60$ g cm$^{-3}$. This agrees with the theory that KBOs and comets are rubble piles and under-dense compared with asteroids, again suggesting a higher proportion of volatile ices and higher porosity. 

\subsection{Rotational properties}

We now compare the distributions of comet and KBO axial-ratios and spin rates separately, to look for similarities in shape and in collisional evolution. KBOs also tend to have low elongations, generally with minimum axial ratios $a/b < 2$. The exceptions are 2003 BF$_{91}$ and 2001 QG$_{\rm 298}$, which are on the right of the plot with $a/b \ge 2.73$ and 2.86 respectively. 2003 BF$_{91}$ is one of the smallest KBOs known, with an absolute visual magnitude of $H_V = 11.7$, corresponding to a radius of 9 km assuming a 10\% albedo \citep{Trilling+Bernstein06}, or $r = 15$ km assuming a comet like 4\% albedo. This small size and large axial ratio suggests that there could be more elongated objects among the smaller KBOs, which have not been detected due to the extreme faintness of these objects. 2001 QG$_{\rm 298}$ also shows a large photometric variation, which would correspond to a very large elongation for a single $r \approx 120$ km body, however this object is thought to be a contact binary due to the shape of its light-curve \citep{Sheppard+Jewitt04}. 

   \begin{figure}
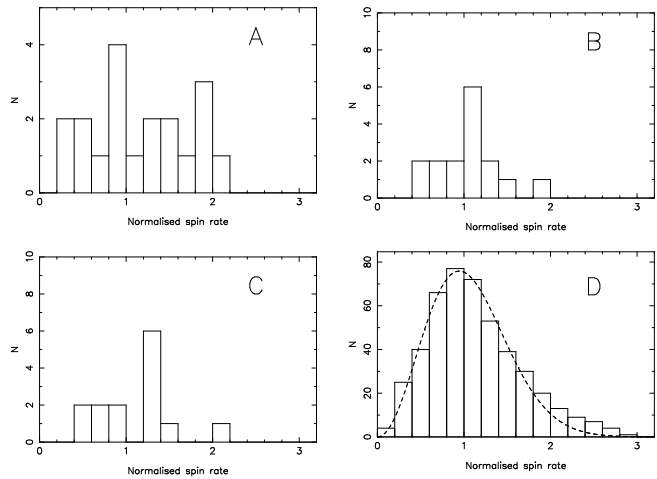

   \centering
   \begin{tabular}{c c}
   \includegraphics[angle=-90,width=0.23\textwidth]{fig17a.ps} &
   \includegraphics[angle=-90,width=0.23\textwidth]{fig17b.ps} \\
   \includegraphics[angle=-90,width=0.23\textwidth]{fig17c.ps} &
   \includegraphics[angle=-90,width=0.23\textwidth]{fig17d.ps} \\
   \end{tabular}
      \caption{Histogram of normalised spin rates ($f/\langle f\rangle$, where $f=1/P_{\rm rot}$) rates for (A) JFCs, (B) KBOs, (C) Centaurs and (D) asteroids. The asteroidal data is reproduced from \citet{Pravec02}, and shows the Maxwellian distribution in spin rates expected for a collisionally evolved system. 
                    }
         \label{Prothist}
   \end{figure}

Figure~\ref{Prothist} shows histograms of spin rates for JFCs, KBOs and Centaurs. We investigate the rotation rate distribution of JFCs to determine whether they are a collisionally relaxed population, or whether torques due to out-gassing substantially alter this distribution. For a collisionally relaxed population, these distributions should show a Maxwellian shape, as seen in large ($r > 20$ km) asteroids (\citealt{Pravec02} -- their fig.~3 is recreated in fig~\ref{Prothist}(D) for comparison). Following \citet{Pravec02}, we plot the rotation frequency $f = 1/P_{\rm rot}$ normalised using the geometric mean $\langle f\rangle$ of each sample, which accounts for different sized bodies. 

The histograms are not obviously Maxwellian in nature, in fact the comet distribution does not significantly differ from a flat distribution; a Kolmogorov-Smirnov test comparing the normalised frequency distribution in fig.~\ref{Prothist}(A) with a flat distribution with the same mean gives a $D$ statistic of 0.12, implying only a 6\% probability that the distribution is not flat. Na\"ively, we expect that torques due to out-gassing from nuclei will act to flatten the distribution from an initial Maxwellian, as torques will speed up or slow down the rotation, although detailed numerical modelling is required to give the theoretical distribution for a cometary population. Given the small sample size, further light-curves are required for cometary nuclei. 


\subsection{Surface colours}

   \begin{figure}
   \centering
   \includegraphics[angle=-90,width=0.47\textwidth]{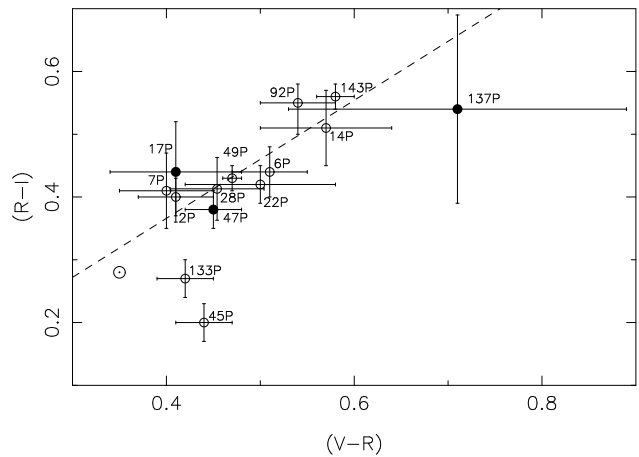}
      \caption{$(R-I)$ against $(V-R)$ for all JFCs with known colours. Filled circles -- this work; open circles -- comets with colours previously determined using the same multi-filter photometry method used here \citep[ and references therein]{Snodgrass05}. The position of the Sun on these axes is marked using the symbol $\odot$.
              }
         \label{colourplot}
   \end{figure}

As described in section~\ref{lc-results}, we take multi-filter images of the comets at different epochs to test for variation in colours which would imply albedo variations over the surface. We find that all of our comets have consistent colours, implying that it is shape, and not large scale albedo features, that is responsible for the observed brightness variations. The measured $(V-R)$ and $(R-I)$ colours are given in table~\ref{results_lightcurve} and plotted in fig.~\ref{colourplot}. These data fit amongst the distribution of known nuclei, with the exception of 137P, which is considerably more red in $(V-R)$ than other comets, although with relatively large error bars which bring it into touch with the others. All three of the comets with new colours presented here fit along the trend found by \citet{Snodgrass05}, which describes an increase in $(R-I)$ with increasing $(V-R)$ indicative of a steadily increasing spectra through these bands, as seen in primitive asteroids and the few cometary nuclei which have been observed spectroscopically \citep{Luu93}. The addition of the new nuclei gives a best fit straight line to the observed trend:
\begin{equation}
(R-I)_{\rm nuc} =m (V-R)_{\rm nuc} + c \left\{ \begin{array}{l}
	m = 0.94 \pm 0.16\\
	c = 0.01 \pm 0.01
	\end{array} \right.
\end{equation}
which is identical to the fit given by \citet{Snodgrass05}, with a very slightly reduced uncertainty on the gradient $m$. In both of these fits, asteroid-comet transition object 133P and comet 45P/Honda-Mrkos-Pajdusakova are rejected as outliers, as they have $(R-I) < (V-R)$, indicating a downturn in the spectra typical of stony asteroids.

   \begin{figure}
   \centering
   \includegraphics[angle=-90,width=0.47\textwidth]{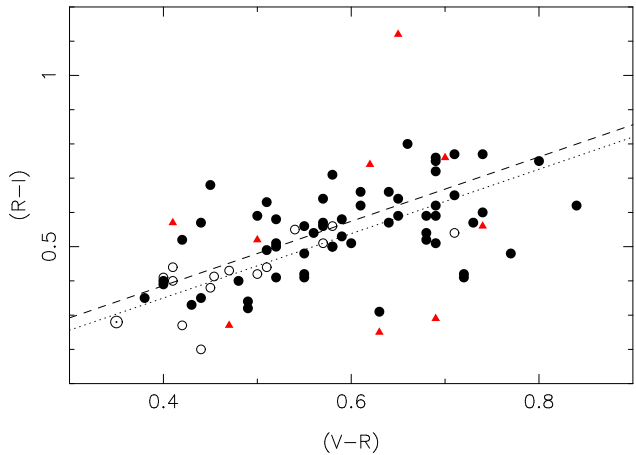}
      \caption{$(R-I)$ against $(V-R)$ for all JFCs, KBOs and Centaurs with known colours. Open circles -- JFCs; filled circles -- KBOs; triangles -- Centaurs. The position of the Sun on these axes is marked using the symbol $\odot$. The dashed line shows our best fit to the comet data, the dotted line is the fit to the KBOs. KBO and Centaur data from \citet{Jewitt+Luu01,Peixinho04}.
              }
         \label{KBOplot}
   \end{figure}

Figure~\ref{KBOplot} is a colour-colour plot showing the data on KBOs (filled circles) and that of JFCs (open circles), with the same trend line over-plotted (KBO data from \citealt{Jewitt+Luu01,Peixinho04}). It is clear that this line provides a reasonable fit to the data, although there is considerable scatter in the KBO results, which is confirmed by a fit to the KBO data which is nearly identical to that for JFCs:
\begin{equation}
(R-I)_{\rm KBO} =m (V-R)_{\rm KBO} + c \left\{ \begin{array}{l}
	m = 0.94 \pm 0.06\\
	c = -0.03 \pm 0.01
	\end{array} \right.
\end{equation}
This is plotted as a dotted line in fig.~\ref{KBOplot}. However, while both populations follow the same trend in colours, it is clear that the KBOs are generally redder than the comet nuclei. The mean colour of JFC nuclei is $\overline{(V-R)}_{\mathrm{nuc}} = 0.46 \pm 0.12$ ($N$ = 23), compared with $\overline{(V-R)}_{\mathrm{KBO}} = 0.59 \pm 0.08$ ($N$ = 62) for the KBO population. 

It has been suggested that `space weathering' produces the very red surfaces seen in KBOs and that cometary activity resurfaces nuclei to give them more neutral colours \citep{Jewitt02}. Our analysis supports this hypothesis; the similar $(V-R)$--$(R-I)$ relations suggesting similar spectra and therefore a common source, while the JFCs tend towards the blue side of the distribution and lack the `ultra-red matter' of KBOs. Under this paradigm, the very red colour we report for 137P may indicate that it is a relatively young comet which has undergone little resurfacing, although we remain cautious of inferring too much from the colour in light of its large uncertainty. We note that the Centaurs (triangles in fig.~\ref{KBOplot}, data also from \citealt{Jewitt+Luu01,Peixinho04}) have a similar spread in $(V-R)$ to the KBOs, but do not appear to show the same trend as the JFCs and KBOs, as they exhibit a very large variation in $(R-I)$. This is somewhat surprising, as Centaurs are thought to be the intermediate dynamical population through which KBOs pass to become JFCs, but there are only nine Centaurs in fig.~\ref{KBOplot} and further data may clarify the situation.

Given the evidence for the loss of extremely red matter due to sublimation and resurfacing, it is important to search for observational signatures of continued surface evolution within the JFC population. For example, \citet{Hughes03} claimed that the size distribution index was shallower for comets with a perihelion distance $q> 2.7$ AU. If reddening of surface material occurs on a time-scale shorter than that for significant orbital change of a JFC, then one might see a correlation with orbital parameters. However, plots of $(V-R)$ against both semi-major axis $a$ and the fraction of orbital period spent at $R_h \le 3$ AU showed no  correlation. Therefore the nucleus colour does not depend on the orbital parameters of the JFC.

A remaining possibility is that the timescale for significant reddening of the surface is shorter than the orbital period. 28P/Neujmin~1 has a large amount of extant photometry (collated by \citealt{Delahodde01} and \citealt{Lamy-chapter}) and is well known as a comet that is essentially dormant throughout much of its orbit, only showing a small amount of activity near perihelion. Taking these colours and plotting them against position in the orbit shows no correlation, it appears that the surface colour of 28P is constant throughout its orbital cycle. Therefore we conclude that significant reddening does not occur on time-scales of less than 20 years and the colours measured for individual comets are not dependant on their orbital position at the time of observation.


\section{Summary}\label{summary}

From time-series $VRI$ photometry of comets 17P, 47P and 137P, acquired during March 2005 using the 3.6m NTT at La Silla, Chile, we place constraints on the sizes, shapes, bulk densities and surface colours of their nuclei. Additionally we present results from snap-shot observations of comets 43P, 44P, 103P and 104P. We find:

\begin{enumerate}
\item Comets 43P and 103P were observed to be active. We place upper limits on the radii of their nuclei at $r_{\rm N} \le 2.4$ km and $\le 4.1$ km respectively. 104P was not detected, allowing us to put a $3\sigma$ upper limit of its radius of $r_{\rm N} \le 0.56$ km, assuming an albedo of 4\%. 

\item The remaining detected comets were stellar in appearance. Surface brightness profiles confirmed that all were effectively inactive. By assuming an albedo of 4\% and a phase coefficient of $\beta = 0.035$ mag deg$^{-1}$, we convert our mean measured magnitudes into radii of $1.96 \pm 0.11$ km for 44P, $1.62 \pm 0.01$ km for 17P, $3.38 \pm 0.01$ km for 47P and $3.58 \pm 0.05$ km for 137P.

\item From our data, we cannot uniquely determine the rotation periods for 17P and 47P, but can constrain the values to a number of solutions. For 17P, the period lies in the region $7.2 \le P_{\rm rot} \le 12.8$ hours, for 47P the period is within $11.2 \le P_{\rm rot} \le 44.0$ hours. We measure minimum axial ratios of $a/b \ge 1.3$ and 1.5 for these two comets, implying minimum densities of between 0.09 and 0.28 g cm$^{-3}$ for 17P, and 0.01 to 0.13 g cm$^{-3}$ for 47P.

\item KBOs have similar rotation periods and elongations to the JFC sample, implying that they have similar densities. Both obey a cut-off in bulk density near 0.6 g cm$^{-3}$. The limited amount of data at present is also consistent with them having the same spin-rate distribution. These physical constraints support the theory that JFCs come from the Kuiper Belt. 

\item We measure colour indices for four nuclei, three of which are typical of other JFCs; $(V-R) = 0.41 \pm 0.07$ and $(R-I) = 0.44 \pm 0.08$ for 17P, $(V-R) = 0.45 \pm 0.03$ and $(R-I) = 0.38 \pm 0.03$ for 47P. For 137P, we measure $(V-R) = 0.71 \pm 0.18$ and $(R-I) = 0.54 \pm 0.15$, which give this comet the reddest nuclear colours measured, albeit with large uncertainties.

\item JFC colours follow a trend of increasing $(R-I)$ with increasing $(V-R)$ that suggests a continually rising spectrum through these bands. This trend also fits KBOs, implying that their spectra have a similar shape. Hence spectrally the JFC population is entirely consistent with being derived from the KBO population but lacking in the extremely red objects. The tendency of nuclei towards the blue end of this trend shows that they are depleted in the `ultra-red matter' that is built up on bodies in the outer solar system by `space weathering'.

\item The optical colours of JFCs show no correlation with orbital parameters. We also note that the extensive photometry of 28P/Neujmin~1 in the literature shows no significant variation in colour as a function of time since it was previously active. Therefore the diversity of JFC nuclei colours is not due to either the orbit or to the time at which the comet was observed.

\end{enumerate}

\section*{acknowledgements}
We thank the staff of ESO's La Silla observatory for their assistance and efforts to maximise the amount of data we could gather during a reduced observing run, and the anonymous referee for helpful comments. IRAF is distributed by the National Optical Astronomy Observatories, which is operated by the Association of Universities for Research in Astronomy, Inc. (AURA) under co-operative agreement with the National Science Foundation.

\bibliography{snodgrass_etal06_NTT}

\setcounter{table}{3}

\begin{table}
\caption{Apparent $R$-band magnitudes for comet 17P.}             
\label{17Ptable}      
\centering                         
\begin{tabular}{@{}ccc@{}}        
\hline                
Date (March 2005) & Airmass & $m_R$ \\    
\hline                        
5.2004 & 1.162  & 22.920$\pm$0.100 \\
5.2019 & 1.157  & 23.058$\pm$0.106 \\
5.2243 & 1.088  & 23.190$\pm$0.118 \\
5.2258 & 1.084  & 22.915$\pm$0.091 \\
5.2273 & 1.080  & 23.122$\pm$0.109 \\
5.2289 & 1.077  & 23.054$\pm$0.103 \\
5.2304 & 1.073  & 22.961$\pm$0.094 \\
5.2318 & 1.070  & 22.910$\pm$0.088 \\
5.2334 & 1.066  & 23.044$\pm$0.103 \\
5.2553 & 1.028  & 22.924$\pm$0.095 \\
5.2588 & 1.024  & 23.040$\pm$0.102 \\
5.2622 & 1.020  & 22.928$\pm$0.094 \\
5.2655 & 1.016  & 22.863$\pm$0.091 \\
5.2690 & 1.013  & 22.972$\pm$0.102 \\
6.1940 & 1.175 & 22.446$\pm$0.108 \\
6.1955 & 1.169 & 22.736$\pm$0.131 \\
6.1970 & 1.163 & 22.932$\pm$0.162 \\
6.1985 & 1.158 & 22.839$\pm$0.147 \\
6.2000 & 1.153 & 22.729$\pm$0.132 \\
6.2015 & 1.147 & 22.820$\pm$0.142 \\
6.2030 & 1.142 & 22.901$\pm$0.151 \\
6.2411 & 1.045 & 22.703$\pm$0.309 \\
6.2458 & 1.038 & 22.956$\pm$0.163 \\
6.2472 & 1.035 & 22.627$\pm$0.147 \\
6.2487 & 1.033 & 22.635$\pm$0.123 \\
6.2501 & 1.031 & 22.828$\pm$0.146 \\
6.2516 & 1.029 & 23.053$\pm$0.197 \\
6.2531 & 1.027 & 22.806$\pm$0.150 \\
6.2546 & 1.025 & 22.836$\pm$0.170 \\
6.3301 & 1.019 & 22.813$\pm$0.145 \\
6.3317 & 1.021 & 22.901$\pm$0.177 \\
6.3352 & 1.025 & 22.635$\pm$0.136 \\
6.3385 & 1.029 & 23.011$\pm$0.194 \\
6.3419 & 1.034 & 22.780$\pm$0.159 \\
6.3656 & 1.080 & 22.868$\pm$0.154 \\
6.3690 & 1.089 & 22.829$\pm$0.148 \\
6.3724 & 1.098 & 22.768$\pm$0.142 \\
6.3758 & 1.107 & 22.876$\pm$0.180 \\
6.3793 & 1.118 & 22.758$\pm$0.139 \\
\hline                                   
\end{tabular}
\end{table}

\begin{table}
\caption{Apparent $R$-band magnitudes for comet 47P.}             
\label{47Ptable}      
\centering                         
\begin{tabular}{@{}ccc@{}}        
\hline 
Date (March 2005) & Airmass & $m_R$ \\    
\hline                        
5.1818 & 1.234 & 21.785$\pm$0.052 \\
5.1832 & 1.229 & 21.713$\pm$0.057 \\
5.1889 & 1.210 & 21.747$\pm$0.058 \\
5.1902 & 1.206 & 21.689$\pm$0.057 \\
5.1915 & 1.202 & 21.693$\pm$0.055 \\
5.1928 & 1.198 & 21.679$\pm$0.055 \\
5.1942 & 1.195 & 21.639$\pm$0.055 \\
5.1955 & 1.191 & 21.727$\pm$0.049 \\
5.1968 & 1.187 & 21.672$\pm$0.052 \\
5.1981 & 1.184 & 21.719$\pm$0.054 \\
5.2361 & 1.123 & 21.623$\pm$0.044 \\
5.2393 & 1.121 & 21.581$\pm$0.043 \\
5.2426 & 1.120 & 21.582$\pm$0.044 \\
5.2457 & 1.119 & 21.562$\pm$0.043 \\
5.2489 & 1.118 & 21.590$\pm$0.044 \\
5.2731 & 1.127 & 21.590$\pm$0.045 \\
5.2744 & 1.128 & 21.546$\pm$0.043 \\
5.2757 & 1.130 & 21.602$\pm$0.045 \\
5.2770 & 1.131 & 21.520$\pm$0.043 \\
6.1171 & 1.638 & 21.643$\pm$0.075 \\
6.1183 & 1.625 & 21.624$\pm$0.081 \\
6.1196 & 1.611 & 21.645$\pm$0.083 \\
6.1209 & 1.598 & 21.732$\pm$0.082 \\
6.1222 & 1.585 & 21.681$\pm$0.080 \\
6.1235 & 1.572 & 21.623$\pm$0.081 \\
6.2091 & 1.154 & 21.443$\pm$0.052 \\
6.2104 & 1.151 & 21.491$\pm$0.053 \\
6.2117 & 1.149 & 21.650$\pm$0.055 \\
6.2130 & 1.147 & 21.448$\pm$0.051 \\
6.2143 & 1.145 & 21.497$\pm$0.049 \\
6.2156 & 1.143 & 21.425$\pm$0.051 \\
6.2200 & 1.137 & 21.400$\pm$0.051 \\
6.2214 & 1.135 & 21.492$\pm$0.054 \\
6.2716 & 1.129 & 21.479$\pm$0.061 \\
6.2747 & 1.132 & 21.482$\pm$0.058 \\
6.2778 & 1.136 & 21.525$\pm$0.061 \\
6.2809 & 1.140 & 21.469$\pm$0.060 \\
6.2839 & 1.144 & 21.504$\pm$0.061 \\
6.3060 & 1.192 & 21.600$\pm$0.065 \\
6.3073 & 1.195 & 21.665$\pm$0.066 \\
6.3086 & 1.199 & 21.624$\pm$0.067 \\
6.3099 & 1.203 & 21.567$\pm$0.066 \\
6.3111 & 1.206 & 21.604$\pm$0.066 \\
6.3125 & 1.210 & 21.644$\pm$0.064 \\
6.3137 & 1.214 & 21.462$\pm$0.062 \\
6.3150 & 1.219 & 21.640$\pm$0.063 \\
6.3440 & 1.345 & 21.897$\pm$0.142 \\
6.3453 & 1.352 & 21.897$\pm$0.134 \\
6.3466 & 1.359 & 21.861$\pm$0.111 \\
6.3492 & 1.375 & 21.912$\pm$0.114 \\
6.3505 & 1.383 & 21.934$\pm$0.135 \\
6.3518 & 1.391 & 21.730$\pm$0.098 \\
6.3531 & 1.399 & 21.773$\pm$0.105 \\
\hline                                   
\end{tabular}
\end{table}

\begin{table}
\caption{Apparent $R$-band magnitudes for comet 137P.}             
\label{137Ptable}      
\centering                         
\begin{tabular}{@{}ccc@{}}        
\hline                 
Date (March 2005) & Airmass & $m_R$ \\    
\hline                        
6.2606 & 1.078  & 22.689$\pm$0.122 \\
6.2625 & 1.075  & 23.016$\pm$0.175 \\
6.2662 & 1.069  & 22.829$\pm$0.148 \\
6.2681 & 1.066  & 22.938$\pm$0.163 \\
6.2860 & 1.048  & 22.869$\pm$0.155 \\
6.2903 & 1.045  & 23.162$\pm$0.192 \\
6.2946 & 1.043  & 22.773$\pm$0.142 \\
6.2988 & 1.041  & 22.930$\pm$0.161 \\
6.3032 & 1.041  & 22.885$\pm$0.139 \\
6.3174 & 1.043  & 22.837$\pm$0.134 \\
6.3194 & 1.044  & 22.860$\pm$0.155 \\
6.3213 & 1.045  & 22.749$\pm$0.127 \\
6.3233 & 1.047  & 22.717$\pm$0.122 \\
6.3252 & 1.048  & 22.862$\pm$0.136 \\
6.3272 & 1.050  & 22.903$\pm$0.141 \\
6.3552 & 1.088  & 22.792$\pm$0.150 \\
6.3571 & 1.092  & 22.774$\pm$0.142 \\
6.3590 & 1.096  & 22.925$\pm$0.167 \\
6.3609 & 1.100  & 22.762$\pm$0.145 \\
6.3629 & 1.104  & 22.708$\pm$0.134 \\
6.3814 & 1.156  & 22.878$\pm$0.186 \\
6.3833 & 1.162  & 23.140$\pm$0.301 \\
\hline                                   
\end{tabular}
\end{table}

\end{document}